\documentclass[%
 reprint,
 superscriptaddress,
 amsmath,amssymb,
 aps,
 pra,
 floatfix,
]{revtex4-2}

\usepackage{graphicx}% Include figure files
\usepackage{dcolumn}% Align table columns on decimal point
\usepackage{bm}% bold math
\usepackage{bbm}
\usepackage{hyperref}
\hypersetup{
    colorlinks=true,
    linkcolor=black,     
    urlcolor=cyan,
    citecolor=blue,
    }
\usepackage{physics}
\usepackage{csquotes}
\usepackage{simpler-wick}
\usepackage{tikz}
\usepackage{accents}
\usepackage{mathtools}
\usepackage[percent]{overpic}

\begin{document}

% \title{Interaction-mediated non-reciprocal dynamics \\ in an exactly solvable open quantum many-body system}% Force line breaks with \\

% \title{Interaction-Mediated Non-Reciprocal Dynamics in an Exactly Solvable Open Quantum System and Beyond}

% \title{Interaction-Mediated Non-Reciprocal Dynamics\\
%from an Exactly Solvable Open Quantum System to Generic Many-Body Systems}

\title{Interaction-Mediated Non-Reciprocal Dynamics
in Open Quantum Systems: \\
From an Exactly Solvable Model to Generic Behavior}

\author{Pietro Borchia}
\affiliation{Faculty of Physics, University of Vienna, Boltzmanngasse 5, 1090 Vienna, Austria
}%
\author{Johannes Knolle}
\affiliation{Technical University of Munich, TUM School of Natural Sciences, Physics Department, Garching, Germany}%
\affiliation{Munich Center for Quantum Science and Technology (MCQST), Schellingstr. 4, 80799 München, Germany}
\affiliation{Blackett Laboratory, Imperial College London, London SW7 2AZ, United Kingdom}
\author{Andreas Nunnenkamp}
\affiliation{Faculty of Physics, University of Vienna, Boltzmanngasse 5, 1090 Vienna, Austria
}%

\date{\today}% It is always \today, today,

\begin{abstract}
Reservoir engineering has emerged as a powerful paradigm to realize non-reciprocal dynamics in open quantum many-body systems. Here, we show that density-density interactions can transfer bath-induced non-reciprocity between different degrees of freedom. Specifically, we investigate a one-dimensional lattice of spin-$\frac{1}{2}$ fermions with all-to-all Hatsugai-Kohmoto interactions in the presence of an engineered reservoir. We establish the exact solvability of the Lindbladian dynamics and show that the interplay between non-reciprocity and interactions qualitatively reshapes the dynamics of excitations. Remarkably, interactions induce directional drift even in spin sectors that are not directly coupled to the reservoir. By analyzing a driven-dissipative Fermi-Hubbard chain, we show that the same mechanism persists for local interactions. The Hatsugai-Kohmoto model thus emerges as a minimal, exactly solvable platform for interaction-mediated non-reciprocal many-body dynamics.

% INPUT: Reservoir engineering has emerged as a powerful paradigm to realize non-reciprocal dynamics in open quantum systems. Here, we show that interactions can mediate and transfer bath-induced non-reciprocity between different degrees of freedom. Specifically, we consider a one-dimensional lattice of spin-$\frac{1}{2}$ fermions with Hatsugai–Kohmoto interactions and establish the exact solvability of the Lindbladian dynamics. We demonstrate that interactions induce directional drift even in spin sectors not directly coupled to the reservoir. Extending perturbatively to driven-dissipative Fermi–Hubbard chains, we find that the same mechanism persists for local interactions. Our results identify interaction-mediated non-reciprocity as a generic phenomenon in open quantum many-body systems.

% We provide compact analytic expressions and diagnostics that quantify directional and interaction-tuned propagation under periodic boundary conditions, establishing a minimal, exactly solvable model system for interaction-driven, non-reciprocal many-body dynamics

\end{abstract}

\maketitle

\section{Introduction}

The study of out-of-equilibrium quantum matter has become a central theme in condensed matter physics. Away from thermal equilibrium, detailed balance can be broken, allowing effective non-reciprocal couplings to emerge \cite{fruchart_vitelli_2026}.
In quantum systems, one possible route to non-reciprocal dynamics is reservoir engineering, where Hamiltonian dynamics is balanced with dissipative processes \cite{PhysRevX.5.021025}. In these cases, the system is coupled to a structured environment and a consistent account of its properties can be achieved within the Lindblad master equation formalism.
While non-reciprocal behavior is typically imprinted directly by the reservoir, it remains an open question to what extent interactions can mediate or redistribute such non-reciprocal effects across different degrees of freedom in a many-body system.

For spatially extended systems, engineered dissipation can be leveraged to realize directional propagation, e.g.~in one-dimensional lattices.
Applying this framework to non-interacting models has allowed to obtain exact analytical results which revealed that systems coupled to engineered reservoirs can host counterintuitive relaxation dynamics \cite{PhysRevLett.123.170401, PhysRevLett.127.070402, PhysRevB.108.064311}, novel steady state properties \cite{PhysRevB.105.064302}, and topological amplification \cite{PhysRevLett.122.143901, wanjura_brunelli_nunnenkamp_2020}. Once interactions are introduced, they can lead to genuine non-equilibrium phase transitions \cite{ashclerk_phases, soares_brunelli_schirò_2025, brighi_nunnenkamp_2025}. In turn, this naturally stimulates further questions regarding the role that interactions can play in reshaping the dynamics in non-reciprocal systems. 

\begin{figure}[t!]
  \centering
  \begin{tikzpicture}
    \node[inner sep=0] (imgA) {%
      \includegraphics[width=\linewidth]{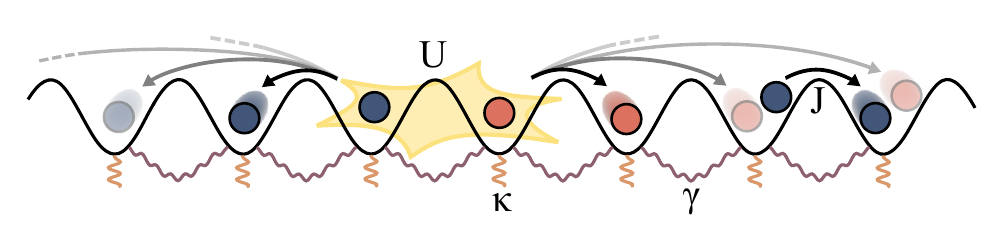}%
    };
    \node[anchor=north west, xshift=0pt, yshift=7pt, font=\scriptsize]
      at (imgA.north west) {(a)};
  \end{tikzpicture}

  \vspace{6pt} % vertical gap between rows
  
  \begin{tikzpicture}
    \node[inner sep=0] (imgB) {%
      \includegraphics[width=\linewidth]{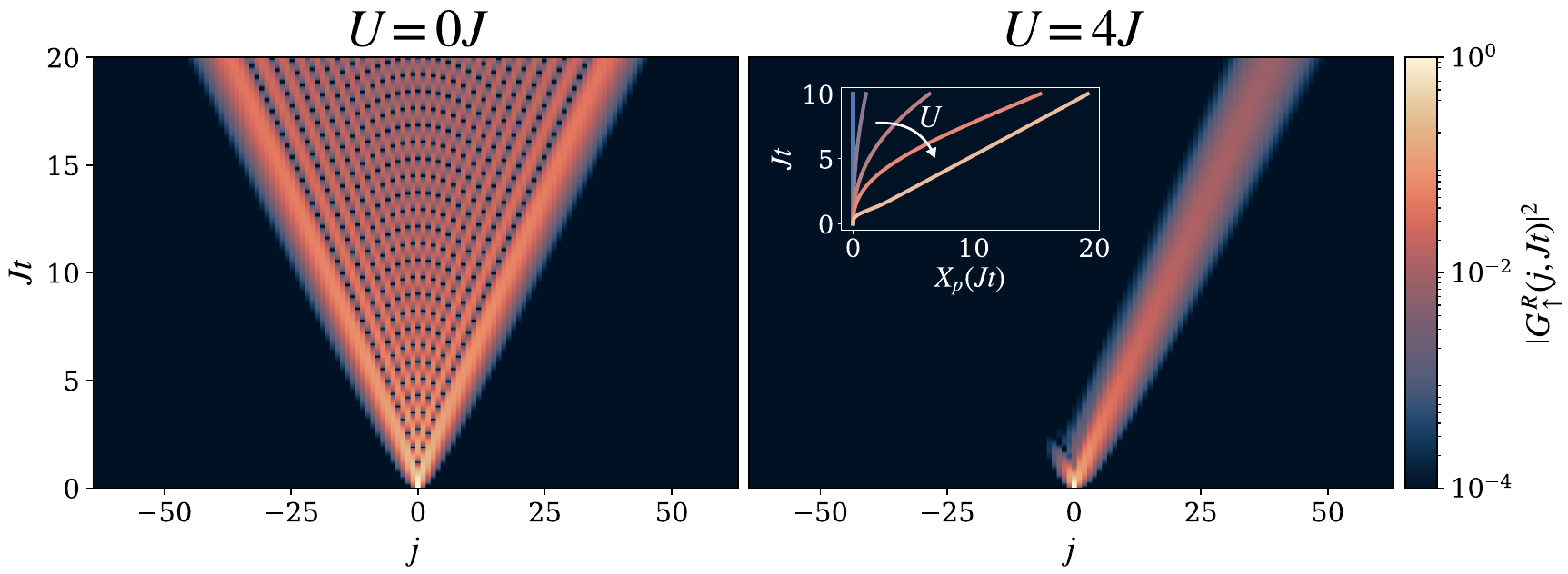}%
    };
    \node[anchor=north west, xshift=-4pt, yshift=7pt, font=\scriptsize]
      at (imgB.north west) {(b)};
    \node[anchor=north east, xshift=4pt, yshift=7pt, font=\scriptsize]
      at (imgB.north) {(c)};
  \end{tikzpicture}
  \caption{(a) Schematic illustration of a one-dimensional lattice of spin-$\frac{1}{2}$ fermions in the presence of two-site loss $\gamma_\downarrow$ and local gain $\kappa_\downarrow$ only on the spin-$\downarrow$ sector, with nearest neighbor hopping strength $J$. Hatsugai-Kohmoto interactions with strength $U$ allow center of mass conserving all-to-all correlated hopping of particles with opposite spin. (b),(c) The real space single particle Green's function $|G_{\uparrow}^R(j,Jt)|^2$ for $U=0$ and $U=4J$, respectively, illustrating non-reciprocal propagation once the two spin sectors are coupled via the interactions in Eq.~(\ref{eq:HK_model}). The inset shows the first moment $X_p(Jt)$  (see Eq.~(\ref{eq:centerofmass})) of the real space retarded Green's function $|G_{\uparrow}^R(j,Jt)|^2$  as the excitation's center of mass for increasing values of the interaction strength $U$ (white arrow). $X_p(Jt)$ serves as a measure of non-reciprocal transport. Increasing $U$ induces a rightward drift. Here, we consider a chain of size $L=128$ where only the spin-$\downarrow$ sector features uniform gain $\kappa_{\downarrow} = 1.5J$ and non-reciprocal loss $\gamma_{\downarrow} = 0.5J$. 
  }
  \label{fig:U-enabled_nonreciprocal}
\end{figure}
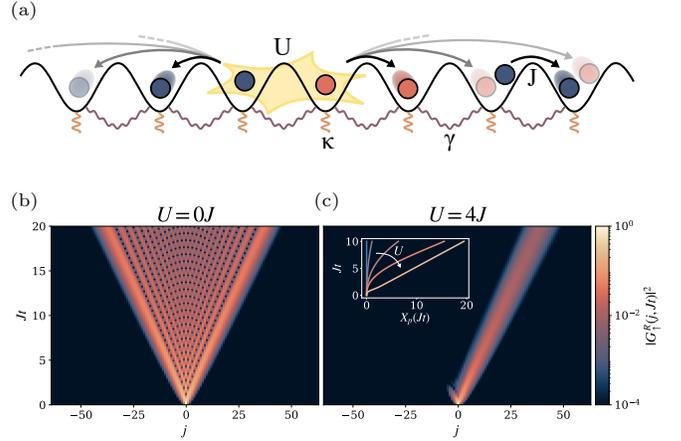
However, adding interactions generally leads to a loss of solvability at the level of the Lindblad master equation. Thus, the majority of works on non-reciprocal many-body systems resort to the no-click approximation, where the generator of the dynamics is given by an effective non-Hermitian Hamiltonian \cite{PhysRevLett.133.076502, PhysRevLett.123.090603, PhysRevB.105.165137, PhysRevB.109.094308, Ashida02072020}. In this limit, the stochastic quantum jump processes induced by the coupling to the environment are completely neglected, a bias that excludes exponentially many quantum trajectories.
To overcome these limitations, several numerical methods have been employed to study non-reciprocal interacting systems under the full-Lindbladian generator and to extract both steady state properties and relaxation dynamics \cite{soares_brunelli_schirò_2025, PhysRevA.110.L020201, PhysRevLett.132.120401}. While these approaches have revealed a rich interplay between interactions and non-reciprocity, they are typically restricted to modest system sizes and offer limited analytical access as well as understanding to the underlying mechanisms.

In this work, we show that density-density interactions can mediate bath-induced non-reciprocal dynamics, enabling directional propagation even in sectors that are not directly coupled to the reservoir. To this end, we introduce a minimal interacting many-body model that remains exactly solvable at the level of the equations of motion.
Concretely, we study a system of spin-$\frac{1}{2}$ fermions coupled to an engineered reservoir in the presence of all-to-all Hatsugai-Kohmoto (HK) interactions. The HK model was originally introduced as a minimal solvable example with an interaction-induced Mott transition~\cite{HK_model_OG}. It has recently gained renewed interest in the context of superconductivity of doped Mott insulators~\cite{phillips2020exact} and for exploring interaction effects in Kondo lattice systems~\cite{zhong2022solvable}, for quantum oscillations~\cite{leeb2023quantum} and topological phenomena~\cite{meng2019unpaired,zhao2023failure}.
Here, we show that the block-diagonal Liouvillian structure inherited from the HK interactions allows one to retain full Lindbladian dynamics while keeping the problem exactly solvable. 

While the steady state is not modified by the interactions, the dynamical response to a linear perturbation is. The problem reduces to an effective non-Hermitian two-mode problem, where interaction-induced hybridization between coherent and dissipative channels enables directional propagation of single-particle excitations even in spin sectors that are not directly coupled to the reservoir, see Fig.~\ref{fig:U-enabled_nonreciprocal}(a). We provide an intuitive picture in terms of the spectral function where the onset of interaction-enabled non-reciprocity is signaled by asymmetric redistribution of spectral weight between different spectral branches and by momentum-selective narrowing of linewidths. Beyond two-time correlators, a closed hierarchy for the off-diagonal elements of the correlation matrix allows us to obtain the exact relaxation of the real space density from an arbitrary initial state. Here, an initially localized particle acquires a directional drift in the presence of interactions, showing that HK interactions can mediate the non-reciprocal character of the reservoir couplings even at the level of density dynamics. 

Finally, we show that this phenomenology is not unique to the fine-tuned HK interactions, but persists for generic local interactions. In a driven-dissipative Fermi-Hubbard chain, the retarded self-energy of the particle conserving sector acquires a complex, momentum-selective dressing via scattering of dissipative particle-hole excitations. Remarkably, this mechanism connects directly to the HK model, providing a minimal, exactly solvable platform for non-reciprocal many-body dynamics.

\section{Model}
We study spin-$\frac{1}{2}$ fermions in a one-dimensional chain under periodic boundary conditions (PBC) with Hatsugai-Kohmoto interactions \cite{HK_model_OG} of strength $U$
\begin{align}\label{eq:HK_model}
    \hat{H} = &-J \sum_{j\sigma}( \hat{c}^\dagger_{j\sigma}\hat{c}_{j+1\sigma} + \text{H.c.}) \notag \\
    &+ \frac{U}{L} \sum_{j_1 \dots j_4} \delta_{j_1+j_3,j_2+j_4} \hat{c}_{j_1\uparrow}^\dagger \hat{c}_{j_2\uparrow} \hat{c}^\dagger_{j_3\downarrow} \hat{c}_{j_4\downarrow},
\end{align}
where $J$ is the nearest-neighbor hopping amplitude and $\hat{c}^\dagger_{j\sigma}\;(\hat{c}_{j\sigma})$ creates (annihilates) a fermion with spin $\sigma$ at site $j$. Here, $L$ is the number of lattice sites. The all-to-all character of the interactions, together with the center-of-mass conservation constraint $j_1+j_3=j_2+j_4$, makes the model diagonal in momentum space
\begin{equation} \label{eq:Hamiltonian}
    \hat{H} = \sum_{k \sigma} \varepsilon_k  \hat{n}_{k\sigma} + U \sum_k\hat{n}_{k\uparrow} \hat{n}_{k\downarrow},
\end{equation}
where $k$ is the quasi-momentum index summed over the Brillouin zone $[-\pi, \pi)$, $\hat{n}_{k\sigma}=\hat{c}^\dagger_{k\sigma}\hat{c}_{k\sigma}$ is the particle number operator and  $\varepsilon_k = -2J\cos k$ is the dispersion relation. Notably, the model features density-density interactions which are diagonal in reciprocal space. Thus, since $[\hat{n}_{k\sigma}, \hat{n}_{k'\sigma'}]=0$, the interaction and kinetic terms of the Hamiltonian commute, and the Hamiltonian is block diagonal in momentum $k$. 

Here, we are interested in investigating the dynamical properties of the system when coupled to engineered reservoirs which introduce incoherent loss and gain. In such cases, the time evolution of the system's density matrix $\rho$ is determined by the Lindblad master equation
\begin{equation}\label{eq:master equation}
    \partial_t\rho = -i[\hat{H}, \rho] + \sum_\alpha\hat{L}_\alpha \rho \hat{L}_\alpha^\dagger - \frac{1}{2}\{\hat{L}_\alpha^\dagger \hat{L}_\alpha, \rho\},
\end{equation}
where $\{\hat{L}_\alpha\}$ is a set of jump operators describing the action of the environment on the system. In the following, we will consider (i) local, uniform gain $\hat{L}_{j\sigma}^{(1)} = \sqrt{\kappa_{\sigma}} \hat{c}^\dagger_{j\sigma}$ and (ii) two-site loss explicitly breaking time reversal-symmetry $\hat{L}_{j\sigma}^{(2)} = \sqrt{\gamma_{\sigma}}(\hat{c}_{j\sigma} + i\hat{c}_{j+1\sigma})$, where $\gamma_\sigma, \kappa_\sigma$ are the corresponding coupling strengths for the respective spin sector, see Fig.~\ref{fig:U-enabled_nonreciprocal}(a) for a schematic depiction of the system. The two-site jump operators introduce a nearest-neighbor dissipative coupling which, in the presence of a finite hopping $J \neq 0$ gives rise to non-reciprocal dynamics~\cite{PhysRevX.5.021025}. 

 Transforming the jump operators (ii) to momentum space makes this statement particularly clear. Indeed, they are equivalent to the set of Lindblad operators acting locally in $k$-space $\hat{L}_{k\sigma}^{(2)} = \sqrt{\gamma_{k\sigma}} \hat{c}_{k\sigma}$ with momentum-dependent rates $\gamma_{k\sigma} = 2 \gamma_{\sigma} (1-\sin k)$ which makes explicit the non-reciprocal character of the dissipative couplings breaking inversion symmetry $\gamma_{k\sigma} \neq \gamma_{-k\sigma}$.  When considering the jump operators above, each momentum mode decouples also at the level of the master equation. Consequently, the full Liouvillian decomposes as a sum of commuting single-momentum generators $\mathcal{L} = \sum_k \mathcal{L}_k$. This specific block-diagonal structure allows us to obtain exact solutions for the equation of motion of key observables, necessary to characterize the system's dynamical behavior. The non-interacting limit $U=0$ has been extensively studied in previous work \cite{PhysRevB.108.064311, PhysRevB.105.064302}, focusing both on steady state properties and relaxation dynamics. 

For the HK model under consideration, it is clear that the static single-particle properties of the steady state are unaffected by the presence of the interactions: since there is no coupling between different $k$-modes the cross-momentum correlations decay to zero, i.e.~the correlation matrix with elements $\langle \hat{c}_{k\sigma}^\dagger \hat{c}_{q\sigma} \rangle$ is diagonal in the long-time limit. However, the presence of a finite interaction strength $U$ has marked effects on the momentum response of the system in its steady state to a weak linear perturbation, and on the relaxation dynamics of the nonequilibrium population.

\section{NESS excitation propagation}\label{sec:NESS excitations}
To characterize the physics of the system's non-equilibrium steady state (NESS) we go beyond equal time observables and focus our attention on two-point single-particle correlation functions. Specifically, we compute the retarded Green's function $G_{\uparrow}^R(k,t) = -i\theta(t) \Tr[\{c_{k\uparrow}(t), c_{k\uparrow}^\dagger(0)\} \rho_\infty]$ that describes the propagation of a $\uparrow$-excitation on top of the steady state $\rho_\infty$, incorporating both the correlation effects generated by the HK interactions and the coupling to the reservoirs. Remarkably, the Green's function can be computed exactly within the Lindblad framework by means of the quantum regression theorem (QRT) \cite{QNoise}. In its fermionic formulation, the QRT leads to the following equation of motion 
\begin{align}\label{eq:EOM_Green_function}
  &\partial_t \langle \hat{c}_{k\uparrow}(t) \hat{c}_{k\uparrow}^\dagger(0) \rangle = i\langle [\hat{H}, \hat{c}_{k\uparrow}](t) \, \hat{c}_{k\uparrow}^\dagger\rangle \notag \\
  &+\sum_\alpha \langle (- \hat{L}_\alpha^\dagger \hat{c}_{k\uparrow} \hat{L}_\alpha - \frac{1}{2}\{\hat{L}_\alpha^\dagger \hat{L}_\alpha, \hat{c}_{k\uparrow}\})(t) \, \hat{c}_{k\uparrow}^\dagger \rangle,
\end{align} 
where the minus sign in front of the quantum jump term $\hat{L}_\alpha^\dagger \cdot \hat{L}_\alpha$ correctly accounts for the particle's statistics, as $\hat{c}_{k\uparrow}$ is odd in the number of fermionic field operators~\cite{PhysRevB.94.155142}. Throughout this section we use the shorthand $\langle \cdot \rangle := \Tr[\cdot \rho_\infty]$. Equation~\eqref{eq:EOM_Green_function} shows that computing the Green’s function reduces to determining the equation of motion of the first moment $\hat{c}_{k\uparrow}$ generated by the single momentum Liouvillian block $\mathcal{L}_k$. For the open HK model, this can be succinctly written as $\partial_t \mathbf{\hat{\Phi}}_k(t) = \mathcal{D}_k \mathbf{\hat{\Phi}}_k(t)$ with $\mathbf{\hat{\Phi}}_k := \begin{pmatrix} \hat{c}_{k\uparrow} ,\;\hat{n}_{k\downarrow}\hat{c}_{k\uparrow} \end{pmatrix}^\intercal$,
where $\mathcal{D}_k$ is the dynamical matrix 
\begin{equation}\label{eq:D_Matrix}
    \mathcal{D}_k = \begin{pmatrix}
        -i\varepsilon_k - \frac{\Gamma_{k\uparrow}}{2} & -iU \\
        \kappa_{\downarrow} & -i(\varepsilon_k+U) - ({\frac{\Gamma_{k\uparrow}}{2}+\Gamma_{k\downarrow}})
    \end{pmatrix},
\end{equation}
where we have introduced $\Gamma_{k\uparrow}=\kappa_{\uparrow}+\gamma_{k\uparrow}$ and analogously $\Gamma_{k\downarrow}$. The retarded Green's function time evolution can then be expressed in terms of the dynamical matrix $G_{\uparrow}^R(k,t) = -i\theta(t) \left[ e^{\mathcal{D}_k t} \mathbf{s}_k \right]_1$ with $\mathbf{s}_k := \langle \{\mathbf{\hat{\Phi}}_k, \hat{c}_{k\uparrow}^\dagger\} \rangle = \begin{pmatrix} 1 \;\;\langle \hat{n}_{k\downarrow} \rangle \end{pmatrix}^\intercal$,
where $[\cdot]_1$ denotes the first component of the vector inside the brackets (see Supplemental Material \cite{Supplemental_material}). In this form, the dynamics is entirely governed by the spectrum and the biorthogonal projectors of the non-Hermitian matrix $\mathcal{D}_k$. Away from fine-tuned exceptional points, the propagator resolves into two dissipative modes
\begin{equation}\label{eq:GF_two_modes}
    G_{\uparrow}^R(k,t) = -i\theta(t) \sum_{\eta=1,2}z_{\eta}(k) e^{t\lambda_{\eta}(k)},
\end{equation}
with 
\begin{equation}\label{eq:eigenvalues}
    \lambda_{1,2} = -i(\varepsilon_k+\frac{U}{2}) - \frac{1}{2}(\Gamma_{k\uparrow}+\Gamma_{k\downarrow}) \pm \frac{1}{2}\sqrt{(\Gamma_{k\downarrow}+iU)^2-4iU\kappa_{\downarrow}}
\end{equation}
where the complex poles $\lambda_{\eta}(k)$ are the eigenvalues of $\mathcal{D}_k$, with corresponding residues $z_{\eta}(k)$ (derivation in the Supplemental Material \cite{Supplemental_material}). 

From Eq.~\eqref{eq:D_Matrix}, we can directly see how interactions and dissipation reshape the system’s linear response, where a crucial role is played by the off-diagonal terms. This is one of the central findings of this paper. A finite $U$ couples the first moment $\hat{c}_{k\uparrow}$ to the opposite spin sector through the higher-order operator $\hat{n}_{k\downarrow}\hat{c}_{k\uparrow}$. Symmetrically, the incoherent pump $\kappa_{\downarrow}$ in the opposite spin sector provides a source term whose contribution is independent of the instantaneous occupation, reflecting the fact that particle injection occurs even for an empty mode, coupling directly $\hat{n}_{k\downarrow}\hat{c}_{k\uparrow}$ to the first moment. Moreover, a finite gain $\kappa_{\downarrow}$ is necessary to avoid a trivial steady state without $\downarrow$-particles that can interact with the $\uparrow$-excitation. If either $U$ or $\kappa_{\downarrow}$ vanishes, the dynamical matrix becomes triangular and the eigenvalues are simply given by the diagonal elements. Conversely, when both $U$ and $\kappa_{\downarrow}$ are finite the two channels hybridize. 

Physically this means that the coherent propagation of an excitation through the lattice is also dressed by the system-reservoir couplings, and that the interaction directly modifies its decay rate and the associated spectral broadening. 
For an excitation initially localized at site $j=0$ in a translationally invariant chain, its real-space propagation is described by the Fourier transform of Eq.~(\ref{eq:GF_two_modes}) $G_{\uparrow}^R(j,t) = L^{-1} \sum_k e^{ikj} G_{\uparrow}^R(k,t)$.
\begin{figure*}[t]
  \centering
    \raisebox{0.0mm}{\begin{tikzpicture}
    \node[inner sep=0] (imgA) {%
      \includegraphics[width=0.9\linewidth]{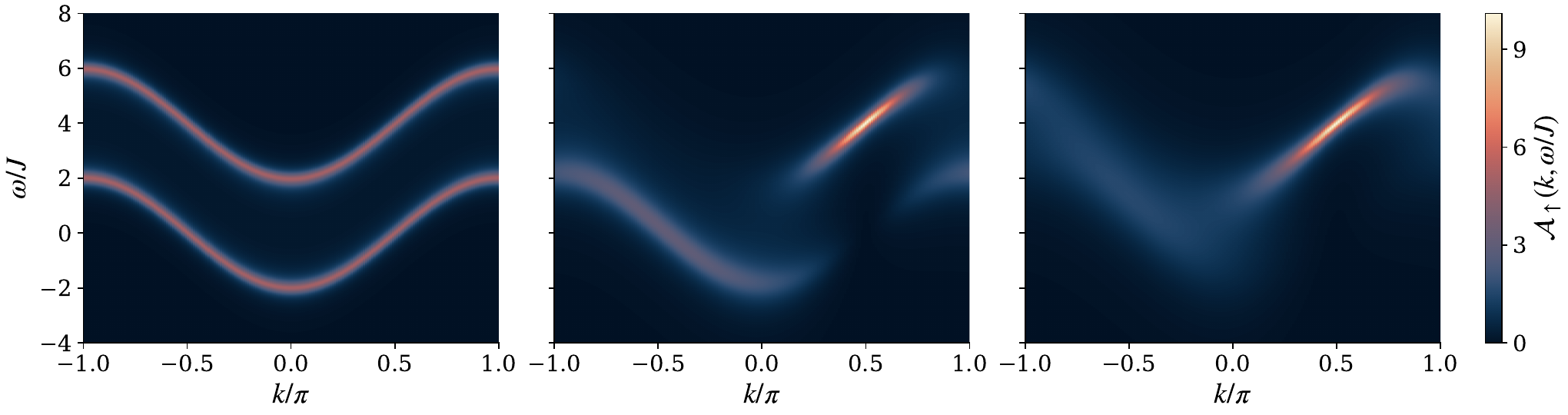}%
    };
    \node[
      anchor=north west,
      xshift=-2pt, yshift=10pt,
      font= \scriptsize
    ] at (imgA.north west) {(a)};
    \node[
      anchor=north,
      xshift=-75pt, yshift=10pt,
      font= \scriptsize
    ] at (imgA.north) {(b)};
    \node[
      anchor=north,
      xshift=65pt, yshift=10pt,
      font= \scriptsize
    ] at (imgA.north) {(c)};
  \end{tikzpicture}
  }
  \caption{(a) Spectral function $\mathcal{A}_\uparrow(k,\omega)$ of the HK model at $U=4J$. For weak, uniform gain and loss on the $\downarrow$-sector ($\gamma_{\downarrow}=\kappa_{\downarrow}=0.2J$), the closed-system two-branch structure survives: dissipation primarily broadens the lines while leaving their centers near $\varepsilon_k$ and $\varepsilon_k+U$. (b) $\mathcal{A}_\uparrow(k,\omega)$ for $U=4J$ with non-reciprocal loss on the $\downarrow$-sector $\gamma_{\downarrow}=0.5J$ and uniform gain $\kappa_{\downarrow}=0.5J$. The $k$-dependent spin-$\downarrow$ occupation reweights the HK branches: near minimal loss ($k_\star\simeq\pi/2$) the upper, interaction-shifted branch $\varepsilon_k+U$ is sharp and dominant, while away from $k_\star$ the lower branch $\varepsilon_k$ takes over and broadens. (c) $\mathcal{A}_\uparrow(k,\omega)$ for $U=4J$ with non-reciprocal loss on the $\downarrow$-sector $\gamma_{\downarrow}=0.5J$ and uniform gain $\kappa_{\downarrow}=1.5J$. As the gain $\kappa_{\downarrow}$ is increased, the low-energy band vanishes and, in the strongly interacting limit $U\gg\kappa_{\downarrow}\gg\gamma_{\downarrow}$, the spectral function approaches that of a non-interacting system with non-reciprocal loss. In Fig (b),(c) uniform loss $\gamma_\uparrow=0.4J$ adds uniform broadening for improved visibility.
  }
  \label{fig:Spectral}
\end{figure*}
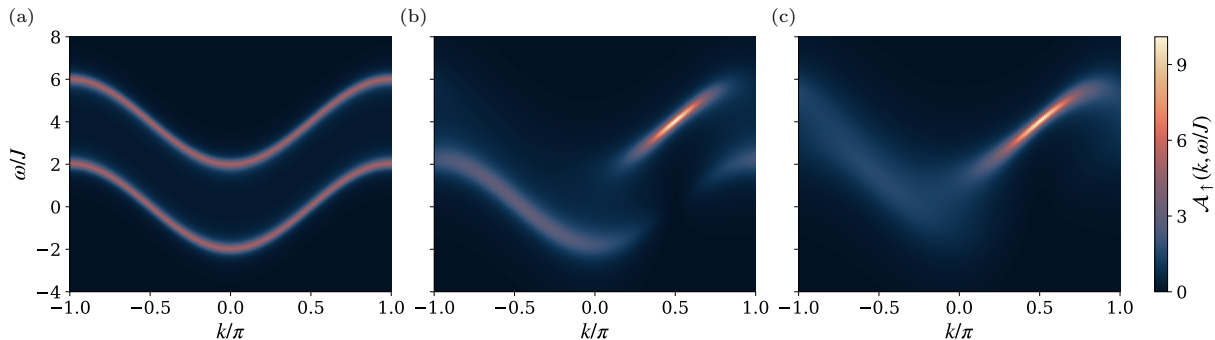
This allows us to immediately gain insight in the interplay between gain, loss and interactions. 

When both baths are uniform in momentum, the dispersion $\varepsilon_k$ is the only $k$-dependent quantity. No interaction-dependent factor then enters the Fourier sum, so the spatial envelope remains the familiar Bessel-function profile of the tight-binding lattice~\cite{Supplemental_material}. The same conclusion holds if non-reciprocity appears only in the $\uparrow$-sector while the $\downarrow$-reservoirs act uniformly: the interaction strength $U$ couples non-trivially only to the $\downarrow$-bath parameters in the dynamical matrix, leaving the spatial envelope governed by the non-interacting form set by the $\uparrow$-sector dissipators. By contrast, once the $\downarrow$-reservoir introduces $k$-dependent gain or loss, the interaction strength enters the Fourier transform, and hence the real-space envelope depends on $U$. In this regime, a finite $U$ yields genuine interaction-mediated non-reciprocal transport. Even when only the $\downarrow$-sector is dissipatively coupled, the interactions imprint directionality onto the $\uparrow$-excitations, which is the second core result of this paper.

Here, we consider a system with uniform gain $\kappa_{\downarrow}=1.5J$ and non-reciprocal loss $\gamma_{\downarrow}=0.5J$ and where the $\uparrow$-sector is not directly coupled to a reservoir. In the non-interacting case $U=0$, the two spin sectors are decoupled and the $\uparrow$-excitation propagates reciprocally, consistent with reservoirs acting directly only on the $\downarrow$-sector [Fig.~\ref{fig:U-enabled_nonreciprocal}(b)]. For finite $U$, interactions mediate the asymmetry of the system-reservoir coupling to the $\uparrow$-sector, suppressing the left-moving light cone; at later times the propagation becomes effectively unidirectional to the right [Fig.~\ref{fig:U-enabled_nonreciprocal}(c)]. The inset in Figure~\ref{fig:U-enabled_nonreciprocal}(c) shows that increasing $U$ drives a clear rightward drift. We quantify this via the first moment of the real-space propagator 
\begin{equation}\label{eq:centerofmass}
    X_p(t)= \frac{\sum_j j|G_{\uparrow}^R(j,t)|^2}{\sum_j |G_{\uparrow}^R(j,t)|^2} ,
\end{equation}
which serves as an effective center of mass and a compact measure of non-reciprocal propagation.  To gain a deeper understanding and a broader picture of the physical consequences of the interplay between dissipation and HK interactions, we now change perspective and examine the spectral function in frequency and momentum.  

\section{Dissipative band structure}
In the previous section we established that a spin sector can acquire non-reciprocal propagation even without being directly coupled to the engineered reservoir, provided it interacts with a dissipative ancillary sector. To connect the real-space directionality to a transparent momentum-frequency picture, we now analyze the spectral function
$\mathcal{A}_\uparrow(k,\omega)=-2\,\mathrm{Im}\,G^R_\uparrow(k,\omega)$,
where $G^R_\uparrow(k,\omega)=\int_{0}^{\infty} dt\, e^{i(\omega+i0^+)t} G^R_\uparrow(k,t)$.
Since for each momentum $k$ the linear response reduces to an effective two-mode non-Hermitian problem [Eq.~\eqref{eq:GF_two_modes}], $G^R_\uparrow(k,\omega)$ is a rational function with two poles. It can therefore be naturally cast in the following form
\begin{equation}
  G^R_\uparrow(k,\omega)= \frac{1}{\omega-\varepsilon_k-\Sigma_\uparrow(k,\omega)},
\end{equation}
which defines an exact retarded self-energy $\Sigma_\uparrow(k,\omega)$ that resums all interaction effects at fixed momentum $k$ enabled by the HK model.
For the open HK chain this self-energy evaluates exactly to
\begin{equation}\label{eq:HK_self}
\Sigma_\uparrow(k,\omega)=U\,n_{k\downarrow}
+U^2\,\frac{n_{k\downarrow}\bigl(1-n_{k\downarrow}\bigr)}
{\omega-\varepsilon_k-U\bigl(1-n_{k\downarrow}\bigr)+i\Gamma_{k\downarrow}},
\end{equation}
with $n_{k\downarrow}=\langle \hat n_{k\downarrow}\rangle$ and where we have set $\Gamma_{k\uparrow}=0$. Equation~\eqref{eq:HK_self} cleanly separates two physical contributions. The first term is a static Hartree shift, which displaces the excitation energy in proportion to the average occupation of the opposite-spin mode. The second term is a dynamical correction controlled by the occupation fluctuations $n_{k\downarrow}(1-n_{k\downarrow})$. It represents the fully resummed dressing of the $\uparrow$-excitation by repeated interaction events with the $\downarrow$-background and acquires an intrinsic dissipative broadening set by $\Gamma_{k\downarrow}$. In the present Lindbladian setting, both $n_{k\downarrow}$ and $\Gamma_{k\downarrow}$ are fixed by the reservoir couplings of the $\downarrow$-sector. Consequently, the environment enters the $\uparrow$-particle response entirely through these quantities. This makes explicit how an asymmetric system-reservoir coupling in the ancillary sector can be transferred to the propagation of $\uparrow$-excitations by interactions.

The closed HK model exhibits two dispersive branches at energies $\varepsilon_k$ and $\varepsilon_k+U$, reflecting whether the opposite-spin mode is empty or occupied. Weak dissipation and gain preserve this structure: the poles remain near these energies while the real parts of the eigenvalues in Eq.~(\ref{eq:eigenvalues}) set finite lifetimes. If the $\downarrow$-baths are momentum independent, then $n_{k\downarrow}$ and $\Gamma_{k\downarrow}$ are $k$-independent, so the interaction-dependent prefactors in Eq.~\eqref{eq:HK_self} do not introduce additional momentum structure beyond $\varepsilon_k$. In this regime the spectral function shows two broadened HK branches with approximately $k$-independent weights, reproducing the symmetric two-peak structure in Fig.~\ref{fig:Spectral}(a).

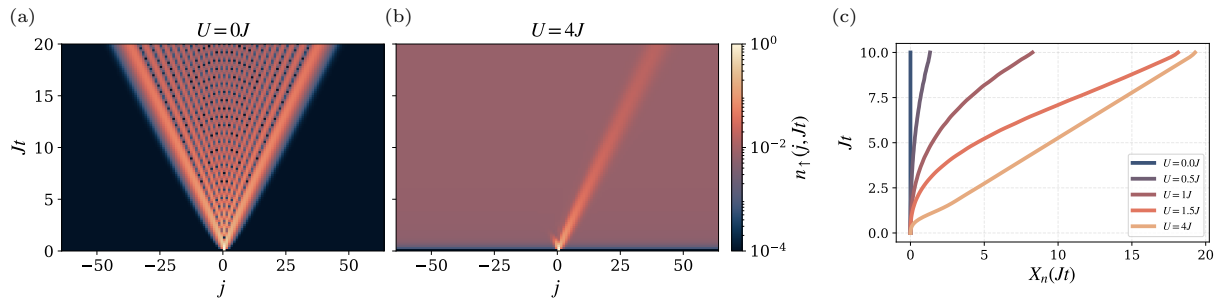
\begin{figure*}[t]
  \centering
  \hspace{0mm}
  % Panel (b) below
  \begin{tikzpicture}
    \node[inner sep=0] (imgB) {%
      \includegraphics[width=0.6\linewidth]{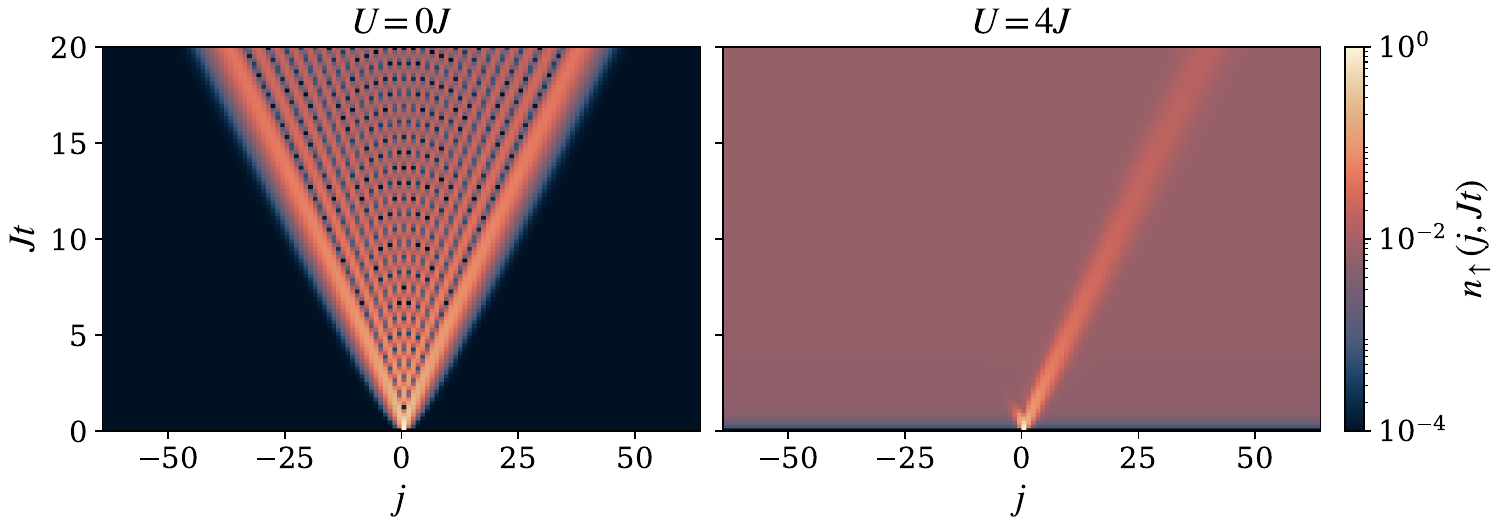}%
    };
    \node[anchor=north west, xshift=-2pt, yshift=7pt, font=\scriptsize] 
      at (imgB.north west) {(a)};
    \node[anchor=north east, xshift=4pt, yshift=7pt, font=\scriptsize]
      at (imgB.north) {(b)};
  \end{tikzpicture}
  % Panel (a) on top
  \begin{tikzpicture}
    \node[inner sep=0] (imgA) {%
      \includegraphics[width=0.3\linewidth]{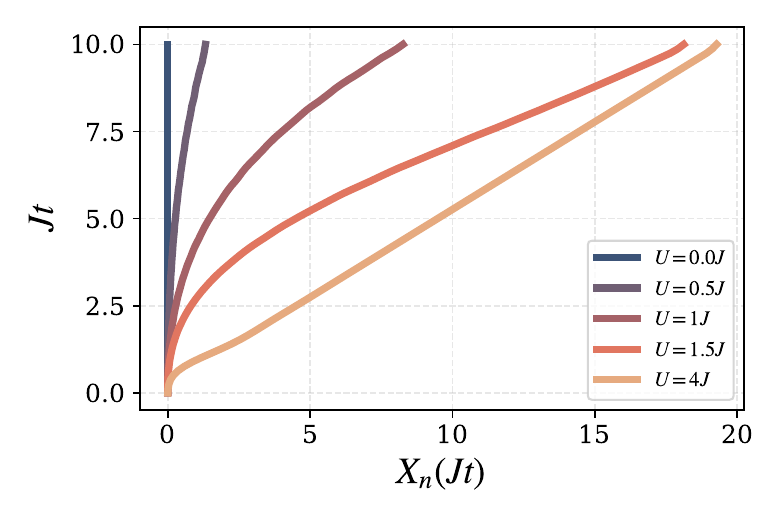}%
    };
    \node[
      anchor=north west,
      xshift=0pt, yshift=11pt,
      font=\scriptsize
    ] at (imgA.north west) {(c)};
  \end{tikzpicture}
  
  \caption{(a) Propagation of the population $n_{\uparrow}(j,Jt)$ through the lattice at $U=0J$ and (b) $U=4J$ displaying interaction-induced non-reciprocal transport of density inhomogeneities. The uniform density background is a signature of the all-to-all nature of the interactions. (c) The nonequilibrium population center $X_n(Jt)$ serves as a measure of non-reciprocal propagation of a $\uparrow$-particle initially localized at $j=0$. As $U$ is increased, the interactions with the opposite spin sector imprint a rightward drift.  In both panels we use a chain of length $L=128$ with only the $\downarrow$-particles subject to uniform gain $\kappa_{\downarrow} = 1.5J$ and non-reciprocal loss $\gamma_{\downarrow} = 0.5J$.}
  \label{fig:density}
\end{figure*}

By contrast, once the $\downarrow$-sector features a momentum-dependent loss $\gamma_{\downarrow}(k)$, the steady-state occupation $n_{k\downarrow}$ becomes strongly $k$-dependent. This momentum selectivity feeds back into Eq.~\eqref{eq:HK_self} in two ways: it redistributes spectral weight between the two HK branches and it produces momentum-dependent linewidths through $\Gamma_{k\downarrow}$. For the jump operators considered here, $\gamma_{\downarrow}(k)$ is minimized at $k_\star=\pi/2$, so in the weak-coupling limit $n_{k_\star\downarrow}\approx 1$ and the spectral weight concentrates on the interaction-shifted branch. Simultaneously, the damping is smallest near $k_\star$, yielding long-lived modes whose group velocity is positive. Therefore, the spectral function becomes dominated by right-moving excitations, as seen in Fig.~\ref{fig:Spectral}(b). We thus identify an unambiguous spectral fingerprint of interaction-mediated non-reciprocity: asymmetric, momentum-selective redistribution of weight between the two HK-branches accompanied by narrowing near the long-lived momenta.

Increasing the gain $\kappa_{\downarrow}$ further drives $n_{k\downarrow}\approx 1$ over a broader portion of the Brillouin zone, progressively suppressing the low-energy branch and concentrating the response on the upper branch [Fig.~\ref{fig:Spectral}(c)]. In the limit $U\gg \kappa_{\downarrow}\gg \gamma_{\downarrow}$, one obtains $z_2(k)\approx 1$ and $\lambda_2(k)\approx -i(\varepsilon_k+U)-\Gamma_{k\uparrow}/2-\gamma_{k\downarrow}$, i.e. up to the constant phase shift $U$ the $\uparrow$-response approaches that of a non-interacting particle subject to an effectively non-reciprocal damping inherited from the opposite spin sector.

\section{Relaxation dynamics}
So far we have discussed how interactions and dissipation reshape the dynamical properties of the steady state by analyzing two-time correlation functions. However, this non-trivial behavior is also displayed by other observable quantities. In particular, we consider the relaxation dynamics of the population in real space $ \hat{n}_{j\uparrow} = \sum_{kq} e^{-i(k-q)j} \hat{c}_{k\uparrow}^\dagger \hat{c}_{q\uparrow}$. Even though interactions do not influence the steady state occupation, they do determine the approach to the steady state of the off-diagonal elements of the correlation matrix $\hat{c}_{k\uparrow}^\dagger \hat{c}_{q\uparrow}$ in momentum space, whose dependence on $U$ is generally non-trivial. Indeed, the equation of motion for the correlation matrix elements is determined by the Lindblad master equation in Eq.~(\ref{eq:master equation}) and for the off-diagonal terms $k \neq q$, it closes upon coupling to higher order terms and is controlled by a $4\times4$ dynamical matrix (see Supplemental Material for the details \cite{Supplemental_material}). 

Specifically, we have $\partial_t \mathbf{\hat{\Psi}}_{kq}(t) = \mathcal{M}_{kq} \mathbf{\hat{\Psi}}_{kq}(t)$ with $\mathbf{\hat{\Psi}}_{kq} := \begin{pmatrix}\hat{c}_{k\uparrow}^\dagger \hat{c}_{q\uparrow} ,\;\hat{c}_{k\uparrow}^\dagger \hat{c}_{q\uparrow} \hat{n}_{k\downarrow},\; \hat{c}_{k\uparrow}^\dagger\hat{c}_{q\uparrow} \hat{n}_{q\downarrow},\; \hat{c}_{k\uparrow}^\dagger\hat{c}_{q\uparrow} \hat{n}_{k\downarrow}\hat{n}_{q\downarrow} \end{pmatrix}^\intercal$,
\begin{widetext}
\begin{equation}
    \mathcal{M}_{kq} = \begin{pmatrix}
            i\Omega_{kq} - \frac{\Gamma_{kq\uparrow}}{2} & iU & -iU & 0 \\
            \kappa_{\downarrow} & i(\Omega_{kq}+U) - \frac{\Gamma_{kq\uparrow}}{2} - \Gamma_{k\downarrow} & 0 & -iU \\
            \kappa_{\downarrow} & 0 & i(\Omega_{kq}-U) - \frac{\Gamma_{kq\uparrow}}{2} - \Gamma_{q\downarrow} & iU \\
            0 & \kappa_{\downarrow} & \kappa_{\downarrow} & i\Omega_{kq} - \frac{\Gamma_{kq\uparrow}}{2}  - \Gamma_{kq\downarrow}
    \end{pmatrix},
\end{equation}
\end{widetext}
where we have defined $\Omega_{kq} = \varepsilon_k - \varepsilon_q$ and $\Gamma_{kq\uparrow}= \Gamma_{k\uparrow}+\Gamma_{q\uparrow}$. Notice that also in this case a pivotal role is played by the interaction strength $U$ and by the gain rate $\kappa_{\downarrow}$ on the opposite spin sector, since both determine the off-diagonal elements of the matrix and hence the non-trivial interplay of interaction and dissipation in the relaxation dynamics, as explained in Sec.~\ref{sec:NESS excitations} for the dynamics of the propagator. 

By solving the eigenvalue problem for the dynamical matrix $\mathcal{M}_{kq}$, we can determine exactly the relaxation dynamics of the correlation matrix for any initial state preparation, and hence the evolution of the population. Notably, the HK interaction does not generate coherence between different $k$-modes and so any spatially uniform initial state will have a trivial evolution. By contrast, inhomogeneous states for which $\langle \hat{c}_{k\uparrow}^\dagger \hat{c}_{q\uparrow} \rangle \neq 0$ exhibit non-trivial dynamics. In such scenarios, interactions reshape the transient evolution of the density profile. 

To illustrate this, we investigate the time evolution of a localized wavefunction prepared in the $\uparrow$-sector, $\langle \hat{n}_{j\uparrow}\rangle = \delta_{j,0}$, on top of a uniform background of $\downarrow$-fermions. Again, to isolate the effect of interactions, we assume the $\uparrow$-sector is effectively closed, while the $\downarrow$-sector is driven by non-reciprocal loss $\gamma_{\downarrow}=0.5J$ and uniform gain $\kappa_{\downarrow}=1.5J$. We initialize the $\downarrow$-sector in its non-equilibrium steady state to provide a constant scattering background. In the non-interacting limit $U=0$, the $\uparrow$-excitation is decoupled from the $\downarrow$-bath; consequently, the density spreads symmetrically [Fig.~\ref{fig:density}(a)]. However, as the interaction strength $U$ is increased, a non-trivial drift emerges [Fig.~\ref{fig:density}(b)].

The resulting dynamics are summarized in Fig.~\ref{fig:density}(c), where we plot the center of mass of the nonequilibrium population, $ X_n(t)  = \sum_j j \delta n_{j\uparrow}(t) / \sum_j \delta n_{j\uparrow}(t)$, with $\delta n_{j\uparrow}(t) = n_{j\uparrow}(t) - n_{j\uparrow}^{ss}$ where $n_{j\uparrow}^{ss}$ is the particle density of site $j$ in the steady state. The off-diagonal coherence $\hat{c}_{k\uparrow}^\dagger \hat{c}_{q\uparrow}$, which encodes the spatial modulation of the density, is dynamically coupled via $\mathbf{\hat{\Psi}}_{kq}$ to mixed correlators involving the non-reciprocal $\downarrow$-background. This asymmetric coupling biases the relaxation of the inhomogeneous component and imprints a preferred direction of motion onto the $\uparrow$-sector wavepacket, confirming that the non-reciprocity of the environment is effectively transferred to the system via the Hatsugai-Kohmoto interactions. 
This demonstrates that the interplay of interactions and non-reciprocity manifests not only in spectral properties of the steady state but also in the transport of density inhomogeneities.

\section{The general lesson: beyond the exactly solvable model}\label{sec:beyond_exact}
So far, our discussion focused on the open HK chain, where the Liouvillian remains block diagonal in momentum space and the full Lindblad dynamics is exactly solvable. A natural question is how general our finding is and whether the same interaction-induced transfer of non-reciprocity persists once the interactions become local in real space, and therefore mix momenta as is generally the case. In this section we show that the mechanism is in fact generic: a particle-conserving sector can acquire directional propagation solely through interactions with a driven-dissipative background that breaks inversion symmetry.

As a minimal setting beyond the HK model, we consider a 1D Fermi-Hubbard chain with spin-dependent dissipation, where the $\downarrow$-sector is coupled to an engineered reservoir while the $\uparrow$-sector is not
\begin{equation}
\hat{H}=\sum_{k,\sigma}\varepsilon_k\,\hat{n}_{k\sigma}
+\frac{U}{L}\sum_{k_1,k_2,q}\hat{c}^\dagger_{k_1\uparrow}\hat{c}^\dagger_{k_2\downarrow}\hat{c}_{k_2-q\downarrow}\hat{c}_{k_1+q\uparrow},
\end{equation}
with jump operators 
\begin{equation}
\hat{L}^{(1)}_{k\downarrow}=\sqrt{\kappa_\downarrow}\,\hat{c}^\dagger_{k\downarrow},
\qquad
\hat{L}^{(2)}_{k\downarrow}=\sqrt{\gamma_{k\downarrow}}\,\hat{c}_{k\downarrow},
\end{equation}
where $\gamma_{k\downarrow}=2\gamma(1-\sin k)$ is inversion breaking. We probe the linear response of a single $\uparrow$-excitation on top of the steady-state $\rho^\text{ss}=|0\rangle\langle 0|_\uparrow\otimes\rho^{\rm ss}_\downarrow$, and we assume that there is no back action of the $\uparrow$-excitation on the spin-$\downarrow$ steady state. In this setting, $\rho^{\rm ss}_\downarrow$ is Gaussian and the correlation matrix is fully characterized by the momentum occupations $n_{k\downarrow}=\langle \hat{n}_{k\downarrow}\rangle$, which inherit the inversion breaking of $\gamma_{k\downarrow}$.

At first order in $U$ one obtains a Hartree shift proportional to the total $\downarrow$-density $ N_\downarrow  = \sum_k n_{k\downarrow} $. Because this contribution is spatially uniform, it does not incorporate the momentum selectivity of the dissipative couplings and therefore cannot imprint any non-reciprocal propagation onto the $\uparrow$-excitations. Directionality instead appears at second order in $U$, where the $\uparrow$-particles can exchange momentum with spin-$\downarrow$ particle-hole excitations.
\begin{figure}[t]
  \centering
  \hspace{0mm}
  % Panel (b) below
  \begin{tikzpicture}
    \node[inner sep=0] (imgB) {%
      \includegraphics[width=0.9\linewidth]{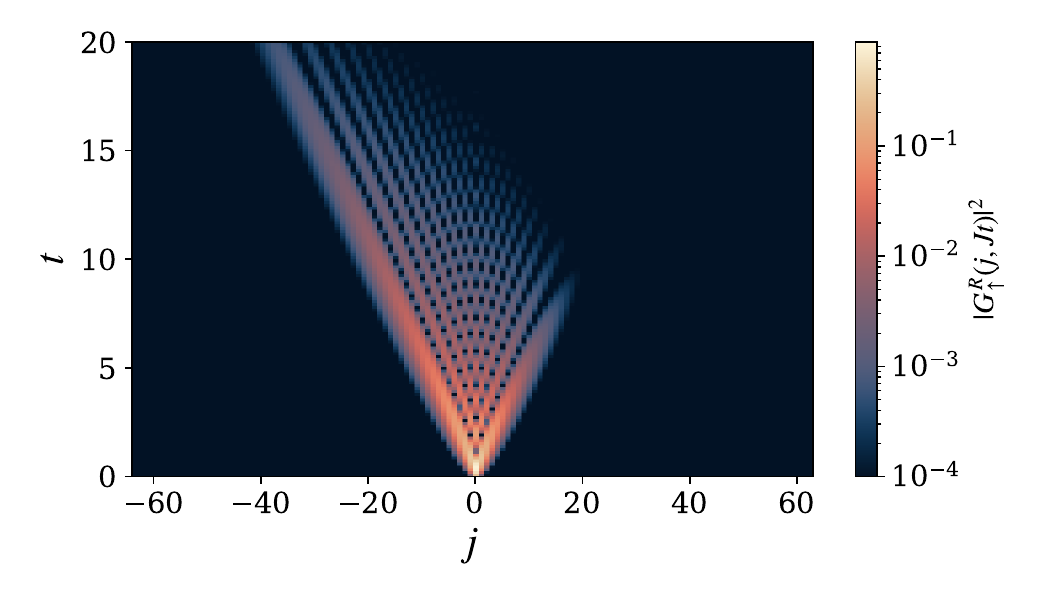}%
    };
    \node[anchor=north west, xshift=-2pt, yshift=7pt, font=\scriptsize] 
      at (imgB.north west) {(a)};
  \end{tikzpicture}
  % Panel (a) on top
  \vspace{6pt} % vertical gap between rows
  
  \begin{tikzpicture}
    \node[inner sep=0] (imgA) {%
      \includegraphics[width=0.9\linewidth]{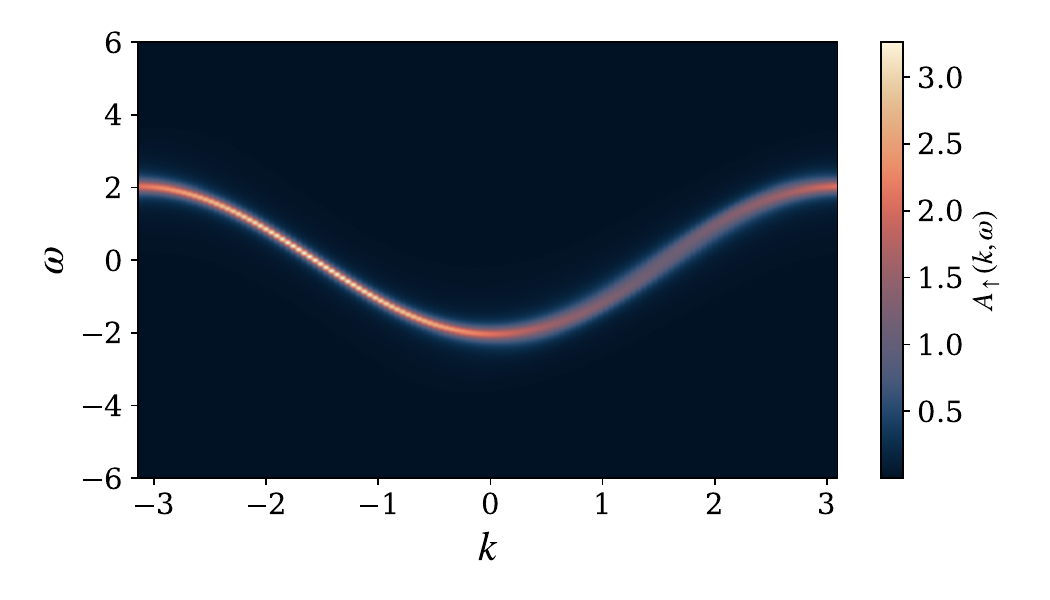}%
    };
    \node[
      anchor=north west,
      xshift=0pt, yshift=7pt,
      font=\scriptsize
    ] at (imgA.north west) {(b)};
  \end{tikzpicture}
  \caption{(a) The real-space single particle Green's function $|G_{\uparrow}^R(j,Jt)|^2$ in the presence of Hubbard interactions, computed via the EOM truncation in Eq.~(\ref{eq:Sigma_Hubbard_U2}). It shows that also for on-site scattering events the excitations of the particle-conserving spin sector acquire a directional character. (b) Momentum-resolved spectral function $\mathcal{A}_\uparrow(k,\omega)$ corresponding to the retarded Green's function in (a). At the level of the momentum-resolved linear response, the presence of the non-reciprocal spin-$\downarrow$ density fluctuations, asymmetrically modifies the spin-$\uparrow$ excitations at different $k$-modes. Here, we consider a translationally invariant 1D Fermi-Hubbard chain of size $L=128$ with $U=2J$, $\kappa_\downarrow=0.2J$ and $\gamma_\downarrow=2J$.}
  \label{fig:FH_GF}
\end{figure}
Carrying out a controlled expansion of the EOM hierarchy at $\mathcal{O}(U^2)$ around the Gaussian spin-$\downarrow$ steady state yields the spin-$\uparrow$ retarded Green's function
\begin{equation}\label{eq:GF_beyond_exact}
G^R_\uparrow(k,\omega)=\frac{1}{\omega-\varepsilon_k-\Sigma_\uparrow(k,\omega)}.
\end{equation}
To leading non-trivial order the self-energy is
\begin{equation}\label{eq:Sigma_Hubbard_U2}
\Sigma_\uparrow(k,\omega)=\frac{U^2}{L^2}\sum_{q,p}\,
\frac{n_{p\downarrow}\bigl(1-n_{p-q\downarrow}\bigr)}
{\omega-\bigl(\varepsilon_{k+q}-\varepsilon_p+\varepsilon_{p-q}\bigr)
+\frac{i}{2}\bigl(\Gamma_{p\downarrow}+\Gamma_{p-q\downarrow}\bigr)} ,
\end{equation}
where $\Gamma_{k\downarrow}= \kappa_\downarrow + \gamma_{k\downarrow}$. Equation~\eqref{eq:Sigma_Hubbard_U2} has a clear interpretation.
The numerator $n_{p\downarrow}(1-n_{p-q\downarrow})$ is the phase-space factor for creating a spin-$\downarrow$ particle-hole pair with momentum transfer $q$, while the denominator contains the corresponding particle-hole energy
$\varepsilon_{k+q}-\varepsilon_p+\varepsilon_{p-q}$. Crucially, the driven-dissipative nature of the $\downarrow$-sector enters not only through the occupations $n_{p\downarrow}$, but also through the intrinsic broadening of the particle-hole continuum, set by the sum of dissipative rates $(\Gamma_{p\downarrow}+\Gamma_{p-q\downarrow})/2$. As a result, inversion breaking in $\gamma_{k\downarrow}$ generically produces an inversion-asymmetric self-energy $\Sigma_\uparrow(k,\omega)\neq \Sigma_\uparrow(-k,\omega)$.

Even though the $\uparrow$-sector is particle conserving and does not couple directly to the reservoir, its self-energy acquires a complex, momentum-selective dressing through repeated scattering of spin-$\downarrow$ particle-hole excitations whose steady-state occupations and lifetimes inherit inversion breaking from coupling to the reservoir. The inversion-asymmetric renormalization implies that the $\uparrow$-excitation momentum-resolved spectral function develops momentum-dependent linewidths and biased spectral-weight redistribution, leading to a suppression of one propagation direction in real space, as shown in Fig.~\ref{fig:FH_GF}. Notice that the same mechanism extends to density-density interactions that depend only on the exchanged momentum, in which case Eq.~(\ref{eq:Sigma_Hubbard_U2}) is generalized by replacing $U \rightarrow U(q)$. The identification of this mechanism is the third central result of this paper.

Remarkably, the HK model is recovered as the special case where the interaction cannot transfer momentum and the particle-hole continuum encoded in the $(p,q)$ sums collapses to a single $k$-mode in momentum space, with the second-order term reducing to a single contribution proportional to the local variance $n_{k\downarrow}(1-n_{k\downarrow})$. Indeed, expanding the exact HK self-energy expression from Eq.~(\ref{eq:HK_self}) at small $U$ yields
\begin{equation}
\Sigma_\uparrow^{\rm HK}(k,\omega)=U n_{k\downarrow}
+\frac{U^2 n_{k\downarrow}(1-n_{k\downarrow})}{\omega-\varepsilon_k+i\Gamma_{k\downarrow}}
+\mathcal{O}(U^3),
\end{equation}
which makes explicit that the open HK chain captures, in closed form, the same underlying mechanism as Eq.~(\ref{eq:Sigma_Hubbard_U2}). 
Therefore, the phenomenology identified in the previous Sections does not rely on the peculiar HK interactions. Inversion-breaking dissipation in an ancillary sector combined with density-density interactions generically induces a momentum-asymmetric self-energy in an otherwise closed sector, thereby producing non-reciprocal propagation of its excitations. The HK model provides a non-perturbative rigorous example where this mechanism can be shown exactly and traced directly to the block structure of the Liouvillian.

\section{Conclusion}
In conclusion, we have investigated the transport properties of a 1D chain of spin-$\tfrac{1}{2}$ fermions with Hatsugai-Kohmoto interactions in the presence of uniform gain and engineered non-reciprocal loss. We have shown that HK interactions can mediate the non-reciprocal features of the coupling to the environment between opposite spin sectors. Thanks to the momentum-diagonal structure of the Liouvillian, the full Lindblad dynamics remains exactly solvable: the single-particle response reduces to an effective non-Hermitian two-mode problem, and the relaxation of off-diagonal correlations is governed by a closed hierarchy involving only four operators. 

At the steady-state level, we derived the retarded Green’s function and spectral function, showing how interaction-induced hybridization with the dissipative channel produces momentum-selective directional propagation, asymmetric redistribution of spectral weight between HK branches, and narrowing of linewidths near long-lived modes favored by the non-reciprocal bath coupling. In real space, we also demonstrated that a localized particle in a spin sector not directly coupled to the reservoir acquires a directional drift when interactions are present, showing that directionality is transmitted also to the transport of density inhomogeneities. 

Beyond the exactly solvable limit, we showed that the same phenomenology is robust. For a driven-dissipative Fermi-Hubbard chain, a systematic expansion to second order in $U$ yields a complex, inversion-asymmetric self-energy for the particle-conserving sector, arising from scattering on particle-hole excitations of the dissipative background. This provides a generic route by which engineered dissipation can imprint directional propagation on sectors that are otherwise decoupled from the reservoir. Remarkably, the HK model can be viewed as an exactly solvable realization of the same mechanism, corresponding to the special case where momentum transfer is suppressed and the particle-hole continuum collapses to a two-channel structure, thereby providing a minimal, exactly solvable platform for interaction-driven non-reciprocal many-body dynamics.
We note that the relationship between the two models was recently discussed in the closed system setting in Refs.~\cite{zhao_hackner_kush_2025, bai_phillips_2025}, suggesting that HK interactions can be used to construct more realistic Hubbard-type models by progressively introducing momentum mixing, opening routes to systematic studies of local interactions in non-reciprocal settings.

Overall, our work motivates several natural directions for future investigation. Starting from the exactly solvable setting studied here, one immediate step is to ask whether momentum-space formulations can be used to construct other interacting Lindbladians that remain analytically tractable. In particular, extending HK-type interactions to systems with orbital degrees of freedom may provide a controlled route to studying topology in interacting open quantum systems \cite{HK_orbitals}. More broadly, this suggests that HK-type interactions may serve as a controlled starting point for understanding interaction-induced effects in a wider range of out-of-equilibrium many-body problems. A second direction is to explore interaction-induced non-reciprocity in genuinely non-integrable settings, where collective excitations emerging beyond the perturbative regime may themselves acquire directional character. In particular, it will be interesting to understand whether a non-reciprocal background can mediate effective interactions that violate action-reaction symmetry in quantum many-body systems.

\begin{acknowledgments}
We thank M.~Buchhold, D.~Malz, Y.~Minoguchi, and C.~Wanjura for fruitful discussions.
This research was funded in whole or in part by the Austrian Science Fund (FWF) [10.55776/COE1, 10.55776/F101200].
For Open Access purposes, the authors have applied a CC-BY public copyright license to any author accepted manuscript version arising from this submission. JK acknowledges support from the Deutsche Forschungsgemeinschaft (DFG, German Research Foundation) under Germany’s Excellence Strategy–EXC–2111–390814868, DFG grants No.~KN1254/1-2, KN1254/2-1, and TRR 360-492547816, as well as the Munich Quantum Valley, which is supported by the Bavarian state government with funds from the Hightech Agenda Bayern Plus.
\end{acknowledgments}

\bibliography{bibsmap}% Produces the bibliography via BibTeX.

%\clearpage
%\setcounter{page}{1}
%\setcounter{section}{0}
%setcounter{subsection}{0}
%\setcounter{figure}{0}
%\setcounter{table}{0}

%\renewcommand{\thesection}{S\arabic{section}}
%\renewcommand{\thefigure}{S\arabic{figure}}
%\renewcommand{\thetable}{S\arabic{table}}

%\include{Supplement}

\end{document}

\title{Supplemental Material for \\ ``Interaction-Mediated Non-Reciprocal Dynamics in Open Quantum Systems: \\ 
From an Exactly Solvable Model to Generic Behavior"}% Force line breaks with \\

\author{Pietro Borchia}
\affiliation{Faculty of Physics, University of Vienna, Boltzmanngasse 5, 1090 Vienna, Austria
}%
\author{Johannes Knolle}
\affiliation{Technical University of Munich, TUM School of Natural Sciences, Physics Department, Garching, Germany}%
\affiliation{Munich Center for Quantum Science and Technology (MCQST), Schellingstr. 4, 80799 München, Germany}
\affiliation{Blackett Laboratory, Imperial College London, London SW7 2AZ, United Kingdom}
\author{Andreas Nunnenkamp}
\affiliation{Faculty of Physics, University of Vienna, Boltzmanngasse 5, 1090 Vienna, Austria
}%

\maketitle
\onecolumngrid
In this Supplemental Material, we provide the derivations underlying the results presented in the main text. In Sec.~\ref{sec:supp_eom}, we derive the closed equations of motion for the fermionic first moment in the open Hatsugai-Kohmoto model and obtain the associated $2\times 2$ dynamical matrix. In Sec.~\ref{sec:supp_green}, we construct the retarded Green's function and the momentum-resolved spectral function. Finally, in Sec.~\ref{sec:supp_beyond}, we outline the perturbative derivation of the order-$U^2$ self-energy for the driven-dissipative Fermi-Hubbard chain beyond the exactly solvable limit.

\section{Equations of motion}\label{sec:supp_eom}
In the main text we showed that the retarded Green's function can be computed exactly thanks to the quantum regression theorem (QRT) \cite{QNoise}. The dynamical matrix governing the time evolution of the two-point correlator can be determined from the EOM of the fermionic field operator $\hat{c}_{k\sigma}$. Here, we derive these equations exactly and show that the hierarchy closes, leading to a $2\times2$ non-Hermitian dynamical matrix.

Recall that a central property of the system's Liouvillian is that each $k$-mode is decoupled. For each, the Hamiltonian is 
\begin{equation}
     H_k = \sum_{\sigma} \varepsilon_k  \hat{n}_{k\sigma} + U \hat{n}_{k\uparrow} \hat{n}_{k\downarrow}
\end{equation}
and the jump operators are $\hat{L}_{k\sigma}^{(1)} = \sqrt{\kappa_\sigma} \hat{c}_{k\sigma}^\dagger$ and $\hat{L}_{k\sigma}^{(2)} = \sqrt{\gamma_{k\sigma}} \hat{c}_{k\sigma}$. We also define $\Gamma_{k\sigma} = \gamma_{k\sigma} + \kappa_{\sigma}$ for later use. The QRT leads to the following equation of motion for the fermionic first moment 
\begin{equation}
    \partial_t \hat{c}_{k\sigma}  = i [\hat{H}_k, \hat{c}_{k\sigma}] +\sum_\alpha (- \hat{L}_\alpha^\dagger \hat{c}_{k\sigma} \hat{L}_\alpha - \frac{1}{2}\{\hat{L}_\alpha^\dagger \hat{L}_\alpha, \hat{c}_{k\sigma}\})(t) \, ,
\end{equation}
where the minus sign in front of the quantum jump term $\hat{L}_\alpha^\dagger \cdot \hat{L}_\alpha$ accounts for the particle's statistics~\cite{PhysRevB.94.155142}. In the derivation of the EOM, we will drop the hat notation $\hat{\cdot}$ and the $k$-mode subscript for the sake of clarity.

\subsection{First moment}\label{subsec:supp_first_moment}
We start by analyzing the Hamiltonian contribution to the dynamics
\begin{align}
    [H_k, c_{\sigma}] &= \varepsilon_k [n_{\sigma}, c_{\sigma}] + U [n_{\sigma} n_{\bar{\sigma}}, c_{\sigma}]  \\
    &= -\varepsilon_k c_{\sigma} - U n_{\bar{\sigma}} c_{\sigma} \nonumber
\end{align}
Notice that the evolution of $\hat{c}_{k\sigma}$ couples to the higher-order term $\hat{n}_{k\bar{\sigma}} \hat{c}_{k\sigma}$, where $\bar{\sigma}$ denotes the opposite spin. 
We now compute the contributions from the coupling to the spin-$\sigma$ reservoirs. The dissipation gives
\begin{align}
    \gamma_{\sigma} (\underbrace{-c_{\sigma}^\dagger c_{\sigma} c_{\sigma}}_{=0} - \frac{1}{2} \{ c_{\sigma}^\dagger c_{\sigma}, c_{\sigma} \}) &= -\frac{1}{2} \gamma_{\sigma} (\underbrace{c_{\sigma}^\dagger c_{\sigma} c_{\sigma}}_{=0} + c_{\sigma} c_{\sigma}^\dagger c_{\sigma}) = -\frac{1}{2} \gamma_{\sigma} c_{\sigma} n_\sigma = \\
    &= -\frac{\gamma_{\sigma}}{2} c_{\sigma} \nonumber
\end{align}
Now, we compute the contributions from the gain 
\begin{equation}
    \kappa_{\sigma} (\underbrace{-c_{\sigma} c_{\sigma} c_{\sigma}^\dagger}_{=0} - \frac{1}{2} \{c_{\sigma} c_{\sigma}^\dagger, c_{\sigma} \}) = -\frac{1}{2} \kappa_{\sigma} (c_{\sigma} c_{\sigma}^\dagger c_{\sigma} + \underbrace{c_{\sigma}^\dagger c_{\sigma} c_{\sigma}}_{=0}) = -\frac{\kappa_{\sigma}}{2} c_{\sigma}
\end{equation}
So, the total contribution from the spin-$\sigma$ channel is
\begin{equation}
    -\frac{1}{2} (\gamma_{\sigma} + \kappa_{\sigma}) c_{\sigma} = -\frac{\Gamma_{\sigma}}{2} c_{\sigma}
\end{equation}
Thus
\begin{equation}
    \partial_t c_{\sigma} = -i\varepsilon_k  c_{\sigma}  - iU n_{\bar{\sigma}} c_{\sigma} - \frac{\Gamma_{\sigma}}{2} c_{\sigma} 
\end{equation}
Notice that there are no contributions from the coupling to spin-$\bar{\sigma}$ reservoirs, as expected for an operator with support only on the $\sigma$-sector.

\subsection{Higher-order term}\label{subsec:supp_higher_order}
We now compute the equation of motion of the higher order term $n_{\bar{\sigma}}c_\sigma$.
\begin{align}
    [H, n_{\bar{\sigma}} c_\sigma] &= \varepsilon_k [n_\sigma, n_{\bar{\sigma}}c_\sigma] + \varepsilon_k [n_{\bar{\sigma}}, n_{\bar{\sigma}}c_\sigma] + U[n_\sigma n_{\bar{\sigma}}, n_{\bar{\sigma}}c_\sigma] = \nonumber \\
    &= \varepsilon_k(n_{\bar{\sigma}}\underbrace{[n_\sigma, c_\sigma]}_{=-c_\sigma} + \underbrace{[n_\sigma, n_{\bar{\sigma}}]}_{=0}c_\sigma + n_{\bar{\sigma}}\underbrace{[n_{\bar{\sigma}},c_\sigma]}_{=0} + \underbrace{[n_{\bar{\sigma}},n_{\bar{\sigma}}]}_{=0}c_\sigma) \nonumber \\
    &\quad + U(n_{\bar{\sigma}}[n_\sigma n_{\bar{\sigma}}, c_\sigma] + \underbrace{[n_\sigma n_{\bar{\sigma}}, n_{\bar{\sigma}}]}_{=0} c_\sigma) = \nonumber \\
    &= -\varepsilon_k n_{\bar{\sigma}}c_\sigma + U n_{\bar{\sigma}} (\underbrace{n_{\sigma}[n_{\bar{\sigma}}, c_\sigma]}_{=0} + \underbrace{[n_\sigma, c_\sigma]}_{=-c_\sigma} n_{\bar{\sigma}}) = \\
    &= -\varepsilon_k n_{\bar{\sigma}}c_\sigma - U n_{\bar{\sigma}} n_{\bar{\sigma}}c_\sigma \nonumber
\end{align}
where $[n_{\bar{\sigma}}, c_\sigma]=0$ and $n_{\bar{\sigma}}^2=n_{\bar{\sigma}}$, thus:
\begin{equation}
[H, n_{\bar{\sigma}} c_\sigma] = -\varepsilon_k  n_{\bar{\sigma}}c_\sigma - U n_{\bar{\sigma}}c_\sigma = -(\varepsilon_k  + U)n_{\bar{\sigma}}c_\sigma
\end{equation}
We compute the contributions from the coupling to the spin-$\sigma$ bath. We analyze the dissipation.
\begin{equation}
    \gamma_\sigma (\underbrace{-c_\sigma^\dagger (n_{\bar{\sigma}}c_\sigma) c_\sigma}_{=0} - \frac{1}{2} \{c_\sigma^\dagger c_\sigma, n_{\bar{\sigma}}c_\sigma\}) = -\frac{\gamma_\sigma}{2}(\underbrace{n_\sigma n_{\bar{\sigma}}c_\sigma}_{=0} + n_{\bar{\sigma}}c_\sigma n_\sigma) = -\frac{\gamma_\sigma}{2} n_{\bar{\sigma}}c_\sigma
\end{equation}
The contributions from the gain are
\begin{equation}
    \kappa_\sigma(\underbrace{-c_\sigma (n_{\bar{\sigma}}c_\sigma) c_\sigma^\dagger}_{=0} - \frac{1}{2}\{c_\sigma c_\sigma^\dagger, n_{\bar{\sigma}}c_\sigma\}) = -\frac{\kappa_\sigma}{2}(c_\sigma c_\sigma^\dagger n_{\bar{\sigma}}c_\sigma + \underbrace{n_{\bar{\sigma}}c_\sigma c_\sigma c_\sigma^\dagger}_{=0}) = -\frac{\kappa_\sigma}{2} c_\sigma n_{\bar{\sigma}}n_\sigma = -\frac{\kappa_\sigma}{2} n_{\bar{\sigma}}c_\sigma
\end{equation}
The total contribution from the spin-$\sigma$ channel is
\begin{equation}
    -\frac{1}{2}\Gamma_\sigma n_{\bar{\sigma}}c_\sigma
\end{equation}
We now compute the contributions from the coupling to the spin-$\bar{\sigma}$ bath. The dissipative channel yields
\begin{equation}
    \gamma_{\bar{\sigma}}(\underbrace{-c_{\bar{\sigma}}^\dagger(n_{\bar{\sigma}}c_\sigma)c_{\bar{\sigma}}}_{=0} - \frac{1}{2}\{c_{\bar{\sigma}}^\dagger c_{\bar{\sigma}}, n_{\bar{\sigma}}c_\sigma\}) = -\frac{\gamma_{\bar{\sigma}}}{2}(n_{\bar{\sigma}}^2 c_\sigma + n_{\bar{\sigma}}c_\sigma n_{\bar{\sigma}}) = -\gamma_{\bar{\sigma}}n_{\bar{\sigma}}c_\sigma
\end{equation}
The contributions from the gain are
\begin{equation}
    \kappa_{\bar{\sigma}}(-c_{\bar{\sigma}}(n_{\bar{\sigma}}c_\sigma)c_{\bar{\sigma}}^\dagger - \frac{1}{2}\{c_{\bar{\sigma}}c_{\bar{\sigma}}^\dagger, n_{\bar{\sigma}}c_\sigma\}) =
\end{equation}
We first focus on the second term:
\begin{equation}
    \{c_{\bar{\sigma}}c_{\bar{\sigma}}^\dagger, n_{\bar{\sigma}}c_\sigma\} = \{(1-n_{\bar{\sigma}}), n_{\bar{\sigma}}c_\sigma\} = 2n_{\bar{\sigma}}c_\sigma - n_{\bar{\sigma}}^2 c_\sigma - n_{\bar{\sigma}}c_\sigma n_{\bar{\sigma}} = 0
\end{equation}
The first term can be simplified
\begin{align}
    c_{\bar{\sigma}} n_{\bar{\sigma}} c_{\bar{\sigma}}^\dagger &= c_{\bar{\sigma}} c_{\bar{\sigma}}^\dagger c_{\bar{\sigma}} c_{\bar{\sigma}}^\dagger = (1-n_{\bar{\sigma}})(1-n_{\bar{\sigma}}) = 1-2n_{\bar{\sigma}}+n_{\bar{\sigma}}^2 = 1-n_{\bar{\sigma}} \\
    -c_{\bar{\sigma}} n_{\bar{\sigma}} c_\sigma c_{\bar{\sigma}}^\dagger &= c_{\bar{\sigma}} n_{\bar{\sigma}} c_{\bar{\sigma}}^\dagger c_\sigma = (1-n_{\bar{\sigma}})c_\sigma = c_\sigma - n_{\bar{\sigma}}c_\sigma
\end{align}
The total contribution from the spin-$\bar{\sigma}$ channel is
\begin{equation}
    -\gamma_{\bar{\sigma}}n_{\bar{\sigma}}c_\sigma + \kappa_{\bar{\sigma}}(c_\sigma - n_{\bar{\sigma}}c_\sigma) = \kappa_{\bar{\sigma}}c_\sigma - (\kappa_{\bar{\sigma}}+\gamma_{\bar{\sigma}})n_{\bar{\sigma}}c_\sigma
\end{equation}
Finally, The EOM for $n_{k\bar{\sigma}}c_{k\sigma}$ is
\begin{equation}
    \partial_t  n_{\bar{\sigma}}c_{\sigma} = -i(\varepsilon_k+U) n_{\bar{\sigma}}c_{\sigma} - (\frac{\Gamma_{\sigma}}{2}+\Gamma_{\bar{\sigma}}) n_{\bar{\sigma}}c_{\sigma}  + \kappa_{\bar{\sigma}} c_{\sigma}
\end{equation}
Therefore, the EOM of the first moment closes by coupling only to one higher-order term.

\subsection{Dynamical matrix}\label{subsec:supp_dynmatrix}
We can now write down the dynamical matrix
\begin{equation}\label{eq:Phi_EOM}
    \partial_t \mathbf{\Phi}_k(t) = \mathcal{D}_k \mathbf{\Phi}_k(t), \qquad \mathbf{\Phi}_k := \begin{pmatrix} c_{k\sigma} \\ n_{k\bar{\sigma}}c_{k\sigma} \end{pmatrix}
\end{equation}
\begin{equation}
    \mathcal{D}_k = \begin{pmatrix}
        -i\varepsilon_k - \frac{\Gamma_{k\sigma}}{2} & -iU \\
        \kappa_{\bar{\sigma}} & -i(\varepsilon_k+U) - ({\frac{\Gamma_{k\sigma}}{2}+\Gamma_{k\bar{\sigma}}})
    \end{pmatrix}
\end{equation}
By solving the eigenvalue problem for the dynamical matrix we can compute the evolution of the first moment and consequently determine the Green's function dynamics. The eigenvalues are
\begin{equation}
    \lambda_{1,2} =  \frac{\text{tr}(\mathcal{D}_k)\pm\sqrt{\Delta}}{2} = -i(\varepsilon_k+\frac{U}{2}) - \frac{1}{2}(\Gamma_{k\sigma}+\Gamma_{k\bar{\sigma}}) \pm \frac{1}{2}\sqrt{(\Gamma_{k\bar{\sigma}}+iU)^2-4iU\kappa_{\bar{\sigma}}}
\end{equation}
where $\Delta = \Tr(\mathcal{D}_k)^2 - 4\det(\mathcal{D}_k)$ clearly displays the interplay between interactions and dissipation.

\section{Green's function}\label{sec:supp_green}
From our previous calculations we saw that $c_{k\sigma}$ dynamics is coupled to the higher order term $Q_{k\sigma} = n_{k\bar{\sigma}}c_{k\sigma}$, which obeys the EOM Eq.~(\ref{eq:Phi_EOM}). We are therefore now set for the calculation of the retarded Green's function
\begin{equation}
    \mathbf{G}_{k\sigma}^R(t) = -i\theta(t) \langle \{\mathbf{\Phi}_k(t), c_{k\sigma}^\dagger(0)\} \rangle = \begin{pmatrix} G_{k,cc}^R(t) \\ G_{k,Qc}^R(t) \end{pmatrix}
\end{equation}
By differentiating we have,
\begin{equation}
    \partial_t \mathbf{G}_{k\sigma}^R(t) = -i\delta(t) \langle \{\mathbf{\Phi}_k(t), c_{k\sigma}^\dagger(0)\} \rangle - i\theta(t) \langle \{\partial_t \mathbf{\Phi}_k(t), c_{k\sigma}^\dagger(0)\} \rangle
\end{equation}
Using the EOM above in the second term we have
\begin{equation}
    (\partial_t - \mathcal{D}_k) \mathbf{G}_{k\sigma}^R(t) = -i\delta(t) \mathbf{s}_k \quad \text{with} \quad \mathbf{s}_k = \langle \{\mathbf{\Phi}_k, c_{k\sigma}^\dagger\} \rangle = \begin{pmatrix} 1 \\ \langle n_{k\bar{\sigma}} \rangle \end{pmatrix}
\end{equation}
where we have used $\{Q_{k\sigma}, c_{k\sigma}^\dagger\} = n_{k\bar{\sigma}}$ for the second component of the $\mathbf{s}_k$ vector. With the causality constraint $G_k^R(t) = 0 \; \; \forall t<0$, the unique solution is
\begin{equation}
    \mathbf{G}_k^R(t) = -i\theta(t) e^{\mathcal{D}_k t} \mathbf{s}_k
\end{equation}
Thus the retarded 1-particle propagator is:
\begin{equation}
    G_{k,cc}^R(t) = -i\theta(t) \langle \{c_{k\sigma}(t), c_{k\sigma}^\dagger(0)\} \rangle = -i\theta(t) \left[ e^{\mathcal{D}_k t} \mathbf{s}_k \right]_1
\end{equation}
where $[\cdot]_1$ denotes the first component of the vector inside the brackets. We can also compute the real-space propagator
\begin{align}
    G_{jl}^R(t) &= \frac{1}{L} \sum_{kq} e^{ikj} G_{kq}^R(t) e^{-iql} \\
    &= \frac{1}{L} \sum_{kq} e^{ik(j-l)} G_{kq}^R(t) e^{i(k-q)l}
\end{align}
Since the system is translationally invariant it means that the result cannot depend on the origin and on $l$ \cite{MB_book}. It follows that $G_{kq} = \delta_{kq}G_k$.
\begin{equation}
    G_{jl}^R(t) = \frac{1}{L} \sum_k e^{ik(j-l)} G_k^R(t)
\end{equation}

\subsection{Spectral function and frequency response}\label{subsec:supp_spectral}
Starting from the Green's function in time, we can derive the response function in frequency space, which provides further insight into the energy distribution of the excitations
\begin{equation}
    \mathcal{A}(k,\omega) = -2\text{Im}[G^R(k,\omega)] \quad \text{where} \quad G^R(k,\omega) = \int_{\mathbb{R}} dt \, e^{i(\omega+i0^+)t} G^R(k,t)
\end{equation}
From previous derivation we obtained
\begin{equation}
    G^R(k,t) = -i\theta(t)[e^{\mathcal{D}_k t}\mathbf{s}_k ]_1 \quad \text{with} \quad \mathbf{s}_k := \begin{pmatrix} 1 \\ \langle n_{k\bar{\sigma}} \rangle \end{pmatrix}, \quad \langle n_{k\bar{\sigma}} \rangle = \frac{\kappa_{\bar{\sigma}}}{\Gamma_{k\bar{\sigma}}}
\end{equation}
Thus we have
\begin{align}
   \mathbf{G}^R(k,\omega) &= \int_{\mathbb{R}} dt \, e^{i(\omega+i0^+)t} \mathbf{G}^R(k,t) = \int_{\mathbb{R}} dt \, e^{i(\omega+i0^+)t} (-i\theta(t)e^{\mathcal{D}_k t}\mathbf{s}_k)  = \\
    &= -i\int_0^\infty dt \, e^{i\omega t - 0^+ t} e^{\mathcal{D}_k t}\mathbf{s}_k  = -i[i\omega-0^++\mathcal{D}_k]^{-1}\mathbf{s}_k  = \\
    &= [(\omega+i0^+)\mathbbm{1}-i\mathcal{D}_k]^{-1}\mathbf{s}_k 
\end{align}
We now define
\begin{equation}
    M_k := (\omega+i0^+)\mathbbm{1}-i\mathcal{D}_k = \begin{pmatrix}
        \omega - \varepsilon_k + i\frac{\Gamma_\sigma}{2} & -U \\
        -i\kappa_{\bar{\sigma}} & \omega-(\varepsilon_k+U)+i(\frac{\Gamma_\sigma}{2}+\Gamma_{\bar{\sigma}})
    \end{pmatrix}
\end{equation}
For a $2\times2$ matrix we can obtain the inverse as
\begin{equation}
    M = \begin{pmatrix} a & b \\ c & d \end{pmatrix} \implies M^{-1} = \frac{1}{\det M} \begin{pmatrix} d & -b \\ -c & a \end{pmatrix}
\end{equation}
The poles $\omega_{1,2}(k)$ that solve $\det M_k=0$ are related to the $D_k$ eigenvalues by $\omega_{1,2}(k) = i\lambda_{1,2}(k)$.
The 1-particle retarded Green's function is given by the (1,1) component
\begin{equation}
    [M_k^{-1}\mathbf{s}_k]_1 = \frac{d-b\langle n_{k\bar{\sigma}} \rangle}{\det(M_k)}
\end{equation}
This gives us the expression
\begin{equation}
    G^R(k,\omega) = \frac{\omega-\varepsilon_k-U(1-\langle n_{k\bar{\sigma}} \rangle)+i(\frac{\Gamma_\sigma}{2}+\Gamma_{\bar{\sigma}})}{[\omega-i\lambda_1(k)][\omega-i\lambda_2(k)]}
\end{equation}
By computing the residues $Z_{1,2}(k)$ we can express the Green's function as:
\begin{equation}
    G^R(k,\omega) = \sum_{j=1,2} \frac{Z_j(k)}{\omega-i\lambda_j(k)}
\end{equation}
Away from fine-tuned exceptional points the Green's function has simple poles and the residues are given by
\begin{equation}
    Z_j(k) = \frac{d-b\langle n_{k\bar{\sigma}} \rangle|_{\omega=\omega_j}}{\partial_\omega \det(M_k)|_{\omega=\omega_j}} = \frac{d-b\langle n_{k\bar{\sigma}} \rangle|_{\omega=\omega_j}}{2\omega-(\omega_1+\omega_2)|_{\omega=\omega_j}}
\end{equation}
So we have:
\begin{align}
    Z_1 &= \frac{U(n_{\bar{\sigma}}^{ss}-\frac{1}{2}) + i\frac{\Gamma_{\bar{\sigma}}}{2}+i\sqrt{\Delta}}{2i\sqrt{\Delta}} \\
    Z_2 &=  \frac{U(n_{\bar{\sigma}}^{ss}-\frac{1}{2}) + i\frac{\Gamma_{\bar{\sigma}}}{2}-i\sqrt{\Delta}}{-2i\sqrt{\Delta}}
\end{align}
Therefore we can compactly write
\begin{equation}
    Z_{1,2} = \frac{1}{2}\left[1 \pm \frac{U(2n_{\bar{\sigma}}^{ss}-1)+i\Gamma_{\bar{\sigma}}}{i\sqrt{\Delta}}\right]
\end{equation}

\section{Beyond the exactly solvable model}\label{sec:supp_beyond}
As a preliminary investigation of the generality of interaction-induced non-reciprocal dynamics, we consider a 1D spin-$\frac{1}{2}$ Fermi-Hubbard chain with incoherent dissipation and drive. The Hubbard interactions are local in real-space and hence lead to all-to-all mixing of $k$-modes across the entire Brillouin zone. In the open setting this model is not generally solvable, and here we compute the momentum-resolved spectral function by a truncation of the EOM. In momentum space the system's Hamiltonian and Lindblad operators are
\begin{equation}
    \mathcal{H}
    =
    \sum_{k,\sigma}\varepsilon_{k\sigma} n_{k\sigma}
    +
    \frac{U}{L}\sum_{k_1,k_2,q}
    c_{k_1\uparrow}^\dagger c_{k_2\downarrow}^\dagger
    c_{k_2-q,\downarrow} c_{k_1+q,\uparrow},
\end{equation}
with jump operators
\begin{equation}
    L_{k\sigma}^{(1)}=\sqrt{\kappa_\sigma}\,c_{k\sigma}^\dagger,
    \qquad
    L_{k\sigma}^{(2)}=\sqrt{\gamma_\sigma(k)}\,c_{k\sigma}.
\end{equation}
We focus on
\begin{equation}
    \gamma_\downarrow(k)=2\gamma[1-\sin k],
    \qquad
    \kappa_\uparrow=\gamma_\uparrow=0,
\end{equation}
so that only the spin-$\downarrow$ sector is driven-dissipative, while the spin-$\uparrow$ sector acts as a probe. Introducing $\rho_{q\sigma}=\sum_k c_{k\sigma}^\dagger c_{k+q,\sigma}$,
the interaction term becomes
\begin{equation}
    \mathcal{H}_{\mathrm{int}}
    =
    \frac{U}{L}\sum_q \rho_{q\uparrow}\rho_{-q\downarrow}.
\end{equation}
Our goal is to derive the effective dynamics of a single \(\uparrow\)-excitation in the non-equilibrium steady state of the \(\downarrow\)-fermions, retaining the leading nontrivial correction in \(U\), namely the order-\(U^2\) self-energy.

\subsubsection{Equation of motion for \texorpdfstring{$c_{k\uparrow}$}{c}}

The Heisenberg equation reads
\begin{equation}
    \partial_t O
    =
    i[\mathcal{H},O]
    +
    \sum_m
    \left(
        \xi\,L_m^\dagger O L_m
        -\frac12\{L_m^\dagger L_m,O\}
    \right),
\end{equation}
with \(\xi=-1\) for odd fermionic operators. Since the jump operators act only on the \(\downarrow\)-sector, the dissipator does not contribute directly to \(c_{k\uparrow}\), and
\begin{equation}
    \partial_t c_{k\uparrow}
    =
    -i\varepsilon_k c_{k\uparrow}
    +
    i[\mathcal{H}_{\mathrm{int}},c_{k\uparrow}].
\end{equation}
Using
\begin{equation}
    [\rho_{q\uparrow},c_{k\uparrow}]
    =
    \sum_p [c_{p\uparrow}^\dagger c_{p+q,\uparrow},c_{k\uparrow}]
    =
    -c_{k+q,\uparrow},
\end{equation}
we obtain the contribution from the interaction term
\begin{equation}
    [\mathcal{H}_{\mathrm{int}},c_{k\uparrow}]
    =
    -\frac{U}{L}\sum_q c_{k+q,\uparrow}\rho_{-q,\downarrow}.
\end{equation}
Therefore, the full time evolution of the fermionic first moment reads
\begin{equation}
    \partial_t c_{k\uparrow}
    =
    -i\varepsilon_k c_{k\uparrow}
    -i\frac{U}{L}\sum_q c_{k+q,\uparrow}\rho_{-q,\downarrow}.
\end{equation}
Expanding the density operator, gives
\begin{equation}
    \partial_t c_{k\uparrow}
    =
    -i\varepsilon_k c_{k\uparrow}
    -i\frac{U}{L}\sum_{q,p}
    c_{k+q,\uparrow}c_{p\downarrow}^\dagger c_{p-q,\downarrow}.
\end{equation}
This suggests introducing  $g_{kqp}:= c_{k+q,\uparrow}c_{p\downarrow}^\dagger c_{p-q,\downarrow}$ so that
\begin{equation}
    \partial_t c_{k\uparrow}
    =
    -i\varepsilon_k c_{k\uparrow}
    -i\frac{U}{L}\sum_{q,p} g_{kqp}.
\end{equation}

\subsubsection{First order in \texorpdfstring{$U$}{U} is trivial}

At first order one would replace
\begin{equation}
    c_{k+q,\uparrow}\rho_{-q,\downarrow}
    \;\to\;
    c_{k+q,\uparrow}\,\langle \rho_{-q,\downarrow}\rangle.
\end{equation}
For the translationally invariant non-interacting steady state of the \(\downarrow\)-fermions,
\begin{equation}
    \langle \rho_{-q,\downarrow}\rangle
    =
    \delta_{q,0}\,\langle N_\downarrow\rangle.
\end{equation}
Thus one only gets the Hartree shift
\begin{equation}
    \partial_t c_{k\uparrow}
    =
    -i\left(
        \varepsilon_k+\frac{U}{L}\langle N_\downarrow\rangle
    \right)c_{k\uparrow}.
\end{equation}
This renormalizes the energy but contains no information about the momentum dependence and non-reciprocity of the driven-dissipative \(\downarrow\)-background, so one must go to order \(U^2\).

\subsubsection{Equation of motion for \texorpdfstring{$g_{kqp}$}{g}}
To extend the hierarchy of equations of motions for the first moment to second order in $U$, we now have to compute the time evolution of the higher-order operator $g_{kqp}$. The free Hamiltonian gives
\begin{equation}
    i[\mathcal{H}_0,g_{kqp}]
    =
    -i(\varepsilon_{k+q}-\varepsilon_p+\varepsilon_{p-q})g_{kqp}.
\end{equation}
Since the jump operators act only on the \(\downarrow\)-sector,
\begin{equation}
    \mathcal{D}(g_{kqp})
    =
    c_{k+q,\uparrow}\,
    \mathcal{D}(c_{p\downarrow}^\dagger c_{p-q,\downarrow}).
\end{equation}
For \(q\neq 0\),
\begin{equation}
    \mathcal{D}(c_{p\downarrow}^\dagger c_{p-q,\downarrow})
    =
    -\frac12\bigl(\Gamma_p+\Gamma_{p-q}\bigr)
    c_{p\downarrow}^\dagger c_{p-q,\downarrow},
\end{equation}
Therefore, the dissipative contribution to the time evolution is
\begin{equation}
    \mathcal{D}(g_{kqp})
    =
    -\frac12\bigl(\Gamma_p+\Gamma_{p-q}\bigr)g_{kqp}.
\end{equation}

For the interaction part, we have
\begin{equation}
    [\mathcal{H}_{\mathrm{int}},g_{kqp}]
    =
    \frac{U}{L}\sum_{q'}
    [\rho_{q'\uparrow}\rho_{-q'\downarrow},
    c_{k+q,\uparrow}c_{p\downarrow}^\dagger c_{p-q,\downarrow}].
\end{equation}
The term involving \(\rho_{q'\uparrow}\) yields
\begin{equation}
    [\rho_{q'\uparrow},g_{kqp}]
    =
    -c_{k+q+q',\uparrow}c_{p\downarrow}^\dagger c_{p-q,\downarrow},
\end{equation}
so that
\begin{equation}
    [\mathcal{H}_{\mathrm{int}},g_{kqp}]
    \simeq
    -\frac{U}{L}\sum_{q'}
    c_{k+q+q',\uparrow}
    c_{p\downarrow}^\dagger c_{p-q,\downarrow}
    \rho_{-q'\downarrow}.
\end{equation}
Here we already discarded the second contribution, proportional to \(\rho_{q'\uparrow}\), since we work in the linear dynamics of a single \(\uparrow\)-excitation around the \(\uparrow\)-vacuum.

Collecting the retained terms,
\begin{equation}
    \partial_t g_{kqp}
    =
    \left[
        -i(\varepsilon_{k+q}-\varepsilon_p+\varepsilon_{p-q})
        -\frac12(\Gamma_p+\Gamma_{p-q})
    \right]g_{kqp}
    -i\frac{U}{L}\sum_{q'}
    c_{k+q+q',\uparrow}
    c_{p\downarrow}^\dagger c_{p-q,\downarrow}\rho_{-q'\downarrow}.
\end{equation}

\subsubsection*{Second-order closure}

We now close the hierarchy by considering the following reference state 
\begin{equation}
    \rho_{\mathrm{ref}}
    =
    |0\rangle\langle 0|_\uparrow
    \otimes
    \rho_{\mathrm{ss},\downarrow},
\end{equation}
where \(\rho_{\mathrm{ss},\downarrow}\) is the non-interacting steady state of the driven-dissipative \(\downarrow\)-fermions, and we neglect the back-action of the single \(\uparrow\)-excitation on this background. Since \(\rho_{\mathrm{ss},\downarrow}\) is Gaussian, Wick's theorem applies.

By translational invariance only the channel \(q'=-q\) contributes, and
\begin{equation}
    c_{k+q+q',\uparrow}
    c_{p\downarrow}^\dagger c_{p-q,\downarrow}\rho_{-q'\downarrow}
    \;\longrightarrow\;
    \delta_{q',-q}\,
    c_{k\uparrow}
    \left\langle
        c_{p\downarrow}^\dagger c_{p-q,\downarrow}\rho_{q\downarrow}
    \right\rangle.
\end{equation}
Using Wick's theorem,
\begin{equation}
    \left\langle
        c_{p\downarrow}^\dagger c_{p-q,\downarrow}\rho_{q\downarrow}
    \right\rangle
    =
    \sum_{p'}
    \left\langle
        c_{p\downarrow}^\dagger c_{p-q,\downarrow}
        c_{p'\downarrow}^\dagger c_{p'+q,\downarrow}
    \right\rangle
    =
    n_p(1-n_{p-q}),
\end{equation}
with $ n_p:=\langle n_{p\downarrow}\rangle_{\mathrm{ss}}$. The closed EOM is then
\begin{equation}
    \partial_t g_{kqp}
    =
    \left[
        -i(\varepsilon_{k+q}-\varepsilon_p+\varepsilon_{p-q})
        -\frac12(\Gamma_p+\Gamma_{p-q})
    \right]g_{kqp}
    -i\frac{U}{L}n_p(1-n_{p-q})\,c_{k\uparrow}.
\end{equation}

\subsubsection*{Frequency-space solution}
Passing to frequency space gives
\begin{equation}
    (\omega-\varepsilon_k)c_{k\uparrow}
    =
    \frac{U}{L}\sum_{q,p} g_{kqp},
\end{equation}
\begin{equation}
    (\omega-\Lambda_{kqp})g_{kqp}
    =
    \frac{U}{L}n_p(1-n_{p-q})\,c_{k\uparrow},
\end{equation}
with $\Lambda_{kqp} := (\varepsilon_{k+q}-\varepsilon_p+\varepsilon_{p-q}) -\frac{i}{2}(\Gamma_p+\Gamma_{p-q})$. Eliminating $g_{kqp}$ we arrive to the expression
\begin{equation}
    (\omega-\varepsilon_k)c_{k\uparrow}
    =
    \frac{U^2}{L^2}
    \sum_{q,p}
    \frac{n_p(1-n_{p-q})}{\omega-\Lambda_{kqp}}
    c_{k\uparrow}.
\end{equation}
This identifies the self-energy
\begin{equation}
    \Sigma_\uparrow(k,\omega)
    =
    \frac{U^2}{L^2}
    \sum_{q,p}
    \frac{n_p(1-n_{p-q})}{\omega-\Lambda_{kqp}},
\end{equation}
and the retarded Green's function
\begin{equation}
    G_\uparrow^R(k,\omega)
    =
    \frac{1}{\omega-\varepsilon_k-\Sigma_\uparrow(k,\omega)}.
\end{equation}
The corresponding virtual process is $k \xrightarrow{\;q\;} k+q \xrightarrow{\;-q\;} k$,
with the non-reciprocal character entering through the distribution $n_p$ and the momentum-dependent linewidth $\Gamma_p$.

\bibliography{bibsmap}